\documentclass[12pt]{article}%
\usepackage{amsmath}
\usepackage{amsfonts}
\usepackage{amssymb}
\usepackage{graphicx}%
\providecommand{\U}[1]{\protect\rule{.1in}{.1in}}
\begin{document}

\title{Density of States in the Magnetic Ground State of the Friedel-Anderson Impurity}
\author{Gerd Bergmann\\Department of Physics\\University of Southern California\\Los Angeles, California 90089-0484\\e-mail: bergmann@usc.edu}
\date{\today}
\maketitle

\begin{abstract}
By applying a magnetic field whose Zeeman energy exceeds the Kondo energy by
an order of magnitude the ground state of the Friedel-Anderson impurity is a
magnetic state. In recent years the author introduced the Friedel Artificially
Inserted Resonance (FAIR) method to investigate impurity properties. Within
this FAIR approach the magnetic ground state is derived. Its full excitation
spectrum and the composition of the excitations is calculated and numerically
evaluated. From the excitation spectrum the electron density of states is
calculated. Majority and minority d-resonances are obtained. The width of the
resonances is about twice as wide as the mean field theory predicts. This
broadening is due to the fact that any change of the occupation of the d-state
in one spin band changes the eigenstates in the opposite spin band and causes
transitions in both spin bands. This broadening reduces the height of the
resonance curve and therefore the density of states by a factor of two. This
yields an intuitive understanding for a previous result of the FAIR approach
that the critical value of the Coulomb interaction for the formation of a
magnetic moment is twice as large as the mean field theory predicts.

 \newpage

\end{abstract}

\section{Introduction}

The properties of magnetic impurities in a metal is one of the most
intensively studied problems in solid state physics. The work of Friedel
\cite{F28} and Anderson \cite{A31} laid the foundation to understand why some
transition-metal impurities form a local magnetic moment while others don't.
Kondo \cite{K8} showed that multiple scattering of conduction electrons by a
magnetic impurity yields a divergent contribution to the resistance in
perturbation theory. Yoshida \cite{Y2} introduced the concept that the (spin
1/2) magnetic impurity forms a singlet state with the conduction electrons and
is non-magnetic at zero temperature. These new insights stimulated a large
body of theoretical and experimental work (see for example \cite{Y2},
\cite{V7}, \cite{S77}, \cite{D44}, \cite{H23}, \cite{M20}, \cite{A36},
\cite{G24}, \cite{C8}, \cite{H20}).

The majority of experimental and theoretical work has focussed on the singlet
Kondo ground state. However, the "magnetic state" of the impurity is of equal
or even greater importance because magnetic impurities are always present,
including in micro-chips and nanostructures, and influence the thermodynamic
and transport properties of the hosts. Since many experiments and almost all
technical applications are \textbf{not }performed at low temperatures the
magnetic impurities are generally far above their Kondo temperature $T_{K}$
and show their full magnetic behavior. The theoretical investigation of the
magnetic state has been explored in much less detail than the Kondo ground
state for spin $1/2$ impurities.

In many cases the Kondo temperature is very low, in the range of liquid helium
temperature. In this case the impurity is in the magnetic state at relatively
low temperature. (The word impurity is in this paper reserved to impurities
which possess - at sufficiently high temperature - a magnetic moment). When
the temperature is several times the Kondo temperature one is sufficiently
above $T_{K}$ to destroy the Kondo ground state. On the other hand one may
expect that the properties of the magnetic state are not yet influenced by the
thermal excitations due to the finite temperature. Therefore a number of
theoretical investigations treat the magnetic state at zero temperature, i.e.
as a magnetic ground state. This approach is probably justified but it leaves
the work always vulnerable to the criticism that there is no magnetic moment
at zero temperature.

Therefore in this paper I prefer to use the effect of a magnetic field on the
Kondo state. A magnetic field which is an order of magnitude larger than
$k_{B}T_{K}/\mu_{B}$ ($\mu_{B}$=Bohr magneton) destroys the Kondo singlet
state as well and yields the magnetic state. Its side effects are that it
changes the energy of the d-states by $\pm\mu_{B}B$ and shifts the conduction
bands by $\pm\mu_{B}B$. The latter yields the Pauli susceptibility but has
otherwise only a negligible effect on the interaction between the impurity and
the conduction electrons because the Fermi level for spin-up and down
electrons readjusts to the same height (as before).

Friedel \cite{F28} and Anderson \cite{A31} derived a criterion for the
instability of the paramagnetic state, i.e. the formation of a magnetic
moment: Take the density of states $N_{d}\left(  \varepsilon_{F}\right)  $ of
the d-resonance at the Fermi energy (in the paramagnetic state) and multiply
it by the Coulomb repulsion energy $U$. If the product $N_{d}U>1$ then a
magnetic moment is formed. Within mean field theory the d-density of states is
given by a Lorentz function%
\[
N_{d,\sigma}\left(  \varepsilon\right)  =\frac{1}{\pi}\frac{\Gamma_{mf}%
}{\left(  \varepsilon-E_{d,\sigma}\right)  ^{2}+\Gamma_{mf}^{2}}%
\]
where $E_{d,\sigma}$ is an effective energy of the d-electrons in the spin-up
or down state, $E_{d,\sigma}=E_{d}+U\left\langle n_{d,-\sigma}\right\rangle $
while $\left\langle n_{d,-\sigma}\right\rangle $ is the average occupation of
the d-electron with the opposite spin) and the resonance width $\Gamma_{mf}$
is given in mean field theory by%
\[
\Gamma_{mf}=\pi\left\vert V_{sd}\right\vert ^{2}N_{s}%
\]
Here $V_{sd}$ is the s-d-hopping matrix element between a conduction electron
and the d-state at the impurity and $N_{s}$ is the density of states of the
conduction electrons. In the mean field theory an occupied $d_{\uparrow}%
^{\dag}$ electron state can only make transitions into $c_{\mathbf{k\uparrow}%
}^{\dag}$-states. (Throughout this paper I express electron states by their
creation operators).

It is well known that the mean field theory has a number of shortcomings.
During the last few years the group of the author has developed a new approach
to the impurity problem, in particular the Friedel-Anderson and the Kondo
impurity. In this \textbf{FAIR} method a \textbf{F}riedel state is
\textbf{A}rtificially \ built from each conduction band and \textbf{I}nserted
as a \textbf{R}esonance state into the conduction or s-band of spin-up and
spin-down electrons. In the appendix a short review of the FAIR solution for
the Friedel impurity is sketched.

The FAIR solution for the magnetic state yields a considerably lower energy
for the "magnetic ground state" and requires a much larger critical Coulomb
interaction to form a magnetic state. This is of some practical importance
since the mean field approximation is used in a number of numerical
spin-density functional theory calculations for the magnetic moment of
impurities in an (s,p) metal host \cite{K46}, \cite{K47}, \cite{M41},
\cite{D28}, \cite{D33}.

In addition to the size of the magnetic moment one would like to know the
density of states in the magnetic state. The answer of the mean field theory
has been discussed above. But there have been a number of suggestions that the
d-resonance is broader than the mean field suggests (see for example Logan
\cite{L57}). The mean field theory decouples the spin-up d-electron from the
spin-down d-electron, but in reality the d-electrons are coupled through the
Coulomb energy. A transition in the $d_{\uparrow}^{\dag}$ electron state
changes the energy and the state of the $d_{\downarrow}^{\dag}$ electron as
well. Therefore it has been suggested in the past that the d-resonances in the
Friedel-Anderson impurity are larger than the mean field theory predicts. A
wider d-resonance in the Friedel-Anderson impurity together with the condition
$N_{d}U>1$ would require a larger Coulomb energy to form a magnetic moment. In
this connection the previous result of the author that the FAIR solution
requires a (two times) larger Coulomb energy to form a magnetic moment would
find a simple physical interpretation.

It is the goal of this paper to calculate the density of states of the
"magnetic ground state" in the FAIR solution and compare it with the mean
field density of states. In section II the theoretical background of the
magnetic state of the Friedel-Anderson impurity is sketched. In section III
electrons and holes are introduced into the magnetic ground state. Their
interactions and excitation energies are derived. In section IV the results of
the numerical calculations are presented. Finally in section V and VI the
results are discussed together with the conclusion. In the appendix A the
basic idea of the FAIR method is sketched.
\[
\]

\section{Theoretical Background}

The simplified Hamiltonian for a magnetic impurity is generally described by
the Friedel-Anderson (FA) Hamiltonian%
\begin{equation}
H_{FA}=%
%TCIMACRO{\tsum _{\sigma}}%
%BeginExpansion
{\textstyle\sum_{\sigma}}
%EndExpansion
\left\{  \sum_{\nu=0}^{N-1}\varepsilon_{\nu}c_{\nu\sigma}^{\dag}c_{\nu\sigma
}+E_{d}d_{\sigma}^{\dag}d_{\sigma}+\sum_{\nu=0}^{N-1}V_{\nu}^{sd}[d_{\sigma
}^{\dag}c_{\nu\sigma}+c_{\nu\sigma}^{\dag}d_{\sigma}]\right\}  +Un_{d\uparrow
}n_{d\downarrow} \label{H_FA}%
\end{equation}
Here the operators $c_{\nu\sigma}^{\dag}$ represent s-electrons, i.e. the
conduction band.

\subsection{The FAIR method}

In the Friedel-Anderson Hamiltonian in equ. (\ref{H_FA}) the d-state for each
spin interacts with every electron in the conduction band. Imagine how much
easier the task would be if the d-electron would interact only with a single
electron state (in each spin band). All other conduction band states would
represent just a background or quasi-vacuum. This is the FAIR approach.

During the last few years the author introduced such a solution to the
Friedel-Anderson impurity problem in which only four electron states, the
spin-up and spin-down d-states $d_{\uparrow}^{\dag}$ and $d_{\downarrow}%
^{\dag}$ and two FAIR states, $a_{0\uparrow}^{\dag}$ and $b_{0\downarrow
}^{\dag}$ interact through the Coulomb and s-d-hopping potential. These states
$a_{0\uparrow}^{\dag}$ and $b_{0\downarrow}^{\dag}$ are composed of the
spin-up and spin-down conduction band states. They are the Friedel
Artificially Inserted Resonance states or FAIR states. The interaction of the
remaining conduction electron states with the d-states is insignificant; they
just yield a background. This yields very good ground-state properties. The
FAIR states are composed of the corresponding conduction bands%
\[%
\begin{array}
[c]{ccc}%
a_{0\uparrow}=%
%TCIMACRO{\tsum _{\nu=0}^{N-1}}%
%BeginExpansion
{\textstyle\sum_{\nu=0}^{N-1}}
%EndExpansion
\alpha_{0}^{\nu}c_{\nu\uparrow} &  & b_{0\downarrow}=%
%TCIMACRO{\tsum _{\nu=0}^{N-1}}%
%BeginExpansion
{\textstyle\sum_{\nu=0}^{N-1}}
%EndExpansion
\beta_{0}^{\nu}c_{\nu\downarrow}%
\end{array}
\]
The remaining $\left(  N-1\right)  $ states in each spin band are constructed
orthogonal to the corresponding FAIR state, orthonormal to each other and
sub-diagonal with respect to the band energy Hamiltonian
\[
H_{0}=\sum_{\nu=0}^{N-1}\varepsilon_{\nu}c_{\nu\sigma}^{\dag}c_{\nu\sigma}%
\]
This yields new bases for the conduction bands $\left\{  a_{i,\uparrow}^{\dag
}\right\}  $ and $\left\{  b_{i,\downarrow}^{\dag}\right\}  $ with $1\leq
i\leq\left(  N-1\right)  $. These new bases are uniquely determined by the two
FAIR states.

Within the new bases the FA-Hamiltonian (\ref{H_FA}) can be expressed as
\[
H_{FA}=H_{0}^{\prime}+H_{1}^{\prime}%
\]
with%
\begin{align*}
H_{0}^{\prime}  &  =H_{0,\uparrow}^{\prime}+H_{0,\downarrow}^{\prime
}+Un_{d\uparrow}n_{d\downarrow}\\
H_{1}^{\prime}  &  =H_{1,\uparrow}^{\prime}+H_{1,\downarrow}^{\prime}%
\end{align*}
where%

\begin{equation}
H_{0,\uparrow}^{\prime}=\sum_{i=1}^{N-1}E_{i}^{\left(  a\right)
}a_{i,\uparrow}^{\dag}a_{i,\uparrow}+E_{0}^{\left(  a\right)  }a_{0,\uparrow
}^{\dag}a_{0,\uparrow}+E_{d}d_{\uparrow}^{\dag}d_{\uparrow}+V_{0}^{\left(
a\right)  sd}\left[  a_{0,\uparrow}^{\dag}d_{\uparrow}+d_{\uparrow}^{\dag
}a_{0,\uparrow}\right]  \label{H_0'}%
\end{equation}%
\begin{equation}
H_{1,\uparrow}^{\prime}=\sum_{i=1}^{N-1}V_{i}^{\left(  a\right)  fr}\left[
a_{0,\uparrow}^{\dag}a_{i,\uparrow}+a_{i,\uparrow}^{\dag}a_{0,\uparrow
}\right]  +\sum_{i=1}^{N-1}V_{i}^{\left(  a\right)  sd}\left[  d_{\uparrow
}^{\dag}a_{i,\uparrow}+a_{i,\uparrow}^{\dag}d_{\uparrow}\right]  \label{H_1'}%
\end{equation}
and the spin-down Hamiltonians are obtained by replacing $\uparrow$ by
$\downarrow$ and the $a_{i}^{\dag}$-states by $b_{i}^{\dag}$-states.

\subsubsection{Nest states}

The Hamiltonian $H_{0}^{\prime}$ is diagonal in the band states $a_{i,\uparrow
}^{\dag}$ and $b_{i,\downarrow}^{\dag}$ for $0<i<N-1$. The only interaction
takes place between the states $a_{0,\uparrow}^{\dag}$, $d_{\uparrow}$,
$b_{0,\downarrow}$ and $d_{\downarrow}$. I call these states the \textbf{nest
states}. The ground state of the Hamiltonian $H_{0}^{\prime}$ is straight
forward. It consists of the coupled state between the nest states and a
partially occupied spin-up and down band. I occupy each spin component with
$N/2$ electrons, putting $n=N/2-1$ electrons into each conduction band states
and one spin-up and one spin-down electron into the nest. This yields the
magnetic ground state as described in equ. (\ref{Psi_MS}).%

\begin{equation}
\Psi_{MS}=\left[  A_{a,b}a_{0\uparrow}^{\dag}b_{0\downarrow}^{\dag}%
+A_{a,d}a_{0\uparrow}^{\dag}d_{\downarrow}^{\dag}+A_{d,b}d_{\uparrow}^{\dag
}b_{0\downarrow}^{\dag}+A_{d,d}d_{\uparrow}^{\dag}d_{\downarrow}^{\dag
}\right]  \left\vert \mathbf{0}_{a,\uparrow}\mathbf{0}_{b,\downarrow
}\right\rangle \label{Psi_MS}%
\end{equation}
where $\left\vert \mathbf{0}_{a,\uparrow}\mathbf{0}_{b,\downarrow
}\right\rangle =\prod_{j=1}^{n-1}a_{j\uparrow}^{\dag}\prod_{j=1}%
^{n-1}b_{j\downarrow}^{\dag}\left\vert \Phi_{0}\right\rangle $ represents a
kind of quasi-vacuum ($n=N/2$).

The calculation of the coefficients $A_{a,b},..$ yields a secular Hamiltonian
$H_{1/1}^{nst}$ which I call the nest-Hamiltonian and which has the form
\[
H_{1/1}^{nst}=\left(
\begin{array}
[c]{cccc}%
E_{0}^{\left(  a\right)  }+E_{0}^{\left(  b\right)  } & V_{b}^{sd} &
V_{a}^{sd} & 0\\
V_{b}^{sd} & E_{0}^{\left(  a\right)  }+E_{d} & 0 & V_{a}^{sd}\\
V_{a}^{sd} & 0 & E_{d}+E_{0}^{\left(  b\right)  } & V_{b}^{sd}\\
0 & V_{a}^{sd} & V_{b}^{sd} & 2E_{d}+U
\end{array}
\right)
\]
Here the abbreviations are used: $V_{a}^{sd}=V_{0}^{\left(  a\right)  sd}$ and
$V_{b}^{sd}=V_{0}^{\left(  b\right)  sd}$ (see equ.(\ref{H_0'})). The
superscript $nst$ stands for nest and the subscript $1/1$ gives the number of
nest electrons in the spin-up and spin-down state. (The energy of the occupied
band states is not included. It yields the same contribution to each
component). $H_{1/1}^{nst}$ has four eigenvalues and eigenstates. The lowest
eigenvalue yields the ground state. The components of the ground state, nest
plus the band states, are shown in Fig.1.

The first order correction to the energy, i.e. the expectation value of
$H_{1}^{\prime}$ is zero. But in addition the second order perturbation of
$H_{1}^{\prime}$ is extremely small. This is demonstrated in appendix B.

It may appear remarkable that the neglect of the interactions between the
d-electron and all the band states $\left\{  a_{j,\uparrow}^{\dag}\right\}  $
and $\left\{  b_{j,\downarrow}^{\dag}\right\}  $ yields a realistic ground
state. But it is not unheard off that one can obtain an excellent ground state
while neglecting a major part of the interaction in the system. The BCS theory
is a good example because it only includes the electron-phonon interaction
between Cooper pairs of time-reversed electrons. The interaction between all
the other electrons is neglected although their number is much larger.

One major part of the numerical calculation is, of course, the optimization of
the two FAIR states $a_{0,\uparrow}^{\dag}$ and $b_{0,\downarrow}^{\dag}$ so
that the expectation value of the energy $E_{00}=\left\langle \Psi
_{MS}\left\vert H_{0}^{\prime}\right\vert \Psi_{MS}\right\rangle $ of the
Hamiltonian $H_{0}^{\prime}$ has a minimum. The optimization procedure is
described at length in previous papers \cite{B151}, \cite{B152}, \cite{B153}
and is taken for granted in this paper and will not be described here. (The
FAIR states are rotated in Hilbert space). Since we don't count the FAIR
states any more as band states the number of band states is reduced by one and
their energy is slightly shifted (by less than the original energy spacing).
The band states enter in the energy $E_{00}$ only through the kinetic (band)
energy of the occupied band states. In a way they just prepare the nest for
the states $\left[  a_{0,\uparrow}^{\dag},d_{\uparrow},b_{0,\downarrow
},d_{\downarrow}\right]  .$
\begin{align*}
&
%TCIMACRO{\FRAME{itbpF}{4.7829in}{1.3117in}{0in}{}{}{corl181_{2}a.eps}%
%width 4.7829in;  height 1.3117in;  depth 0in;  original-width 6.2648in;
%original-height 1.692in;  cropleft "0";  croptop "1";  cropright "1";
%cropbottom "0";  filename 'Fig1.eps';file-properties "XNPEU";}}}%
%BeginExpansion
{\includegraphics[
height=1.3117in,
width=4.7829in
]%
{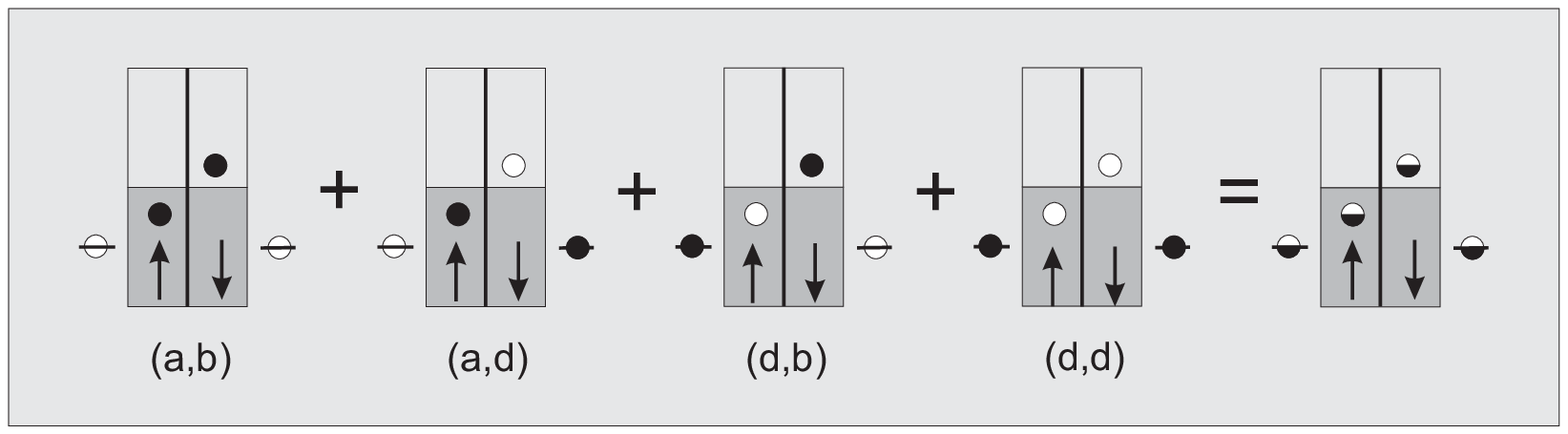}%
}%
%EndExpansion
\\
&
\begin{tabular}
[c]{l}%
Fig.1: The composition of the magnetic state $\Psi_{MS}$ in the nest is
shown.\\
It consists of four Slater states. Each Slater state has a half-full spin-up
and\\
down band, two FAIR states (circles in within the bands) and two d-states\\
(circles on the left and right of the band). Full black circles represent\\
occupied states and light grey represent empty states. The band\ at the
right\\
with the half-filled circles symbolizes the magnetic solution with four\\
Slater states.
\end{tabular}
\end{align*}

In the numerical calculation we will present the results for two examples with
the parameters $U=1.0$, $E_{d}=-0.5$ and $\left\vert V_{sd}^{0}\right\vert
^{2}=0.05$ and $\left\vert V_{sd}^{0}\right\vert ^{2}=0.025$. The smaller
value of the s-d-matrix element permits a better fit of the resulting
resonance curve with a Lorentz curve since the effect of the finite width of
the band is smaller.

\subsubsection{Self-consistent perturbation}

In the construction of the magnetic ground state $\Psi_{MS}$ the Hamiltonian
$H_{1}^{\prime}$ has been completely neglected. Below we will derive the
excitation energies by introducing an additional electron (hole) into an empty
(occupied) states. For this calculation it is important to know whether the
empty state is really empty or whether transitions from the ground state into
the state due to $H_{1}^{\prime}$ have partially occupied this state. (This
problem is well known from the calculation of the electron-phonon mass
enhancement. In the calculation of the electron-phonon self-energy one injects
an electron into an "empty state" $\mathbf{k}$ above the Fermi energy. The
transitions of this electron via the electron-phonon interaction into other
empty states $\mathbf{k}^{\prime}$ contribute to the self-energy $\Sigma$.
However, the state $\mathbf{k}$ was not really empty because transitions from
the ground state into $\mathbf{k}$ already created a finite occupation of
$\mathbf{k}$. One has to correct the self-energy due to these processes).

In the appendix I show that transitions from the ground state into empty band
states (due to $H_{1}^{\prime}$) are practically zero. The interference
between transitions from the d-state and from the corresponding FAIR state
almost perfectly cancel each other. The total weight in all perturbation
states is only of the order of $10^{-4}$ and can be completely neglected.
Therefore the band states are either completely empty or fully occupied.

\section{Calculation of Excitations}

\subsection{Injection of an electron}

In Fig.2 a spin-up electron is injected into one of the empty states
$a_{j\uparrow}^{\dag}$ of the $\left\{  a_{\uparrow}^{\dag}\right\}  $-band.
This yields the Slater states ($\alpha$). The Slater state ($\beta$) is
obtained by injecting an electron into the nest, either into the state
$a_{0\uparrow}^{\dag}$ or $d_{\uparrow}^{\dag}$.
\begin{align*}
&
%TCIMACRO{\FRAME{itbpF}{5.335in}{1.8198in}{0in}{}{}{corl181_{2}d.eps}%
%width 5.335in;  height 1.8198in;  depth 0in;  original-width 5.0668in;
%original-height 1.7111in;  cropleft "0";  croptop "1";  cropright "1";
%cropbottom "0";  filename 'Fig2.eps';file-properties "XNPEU";}}}%
%BeginExpansion
{\includegraphics[
height=1.8198in,
width=5.335in
]%
{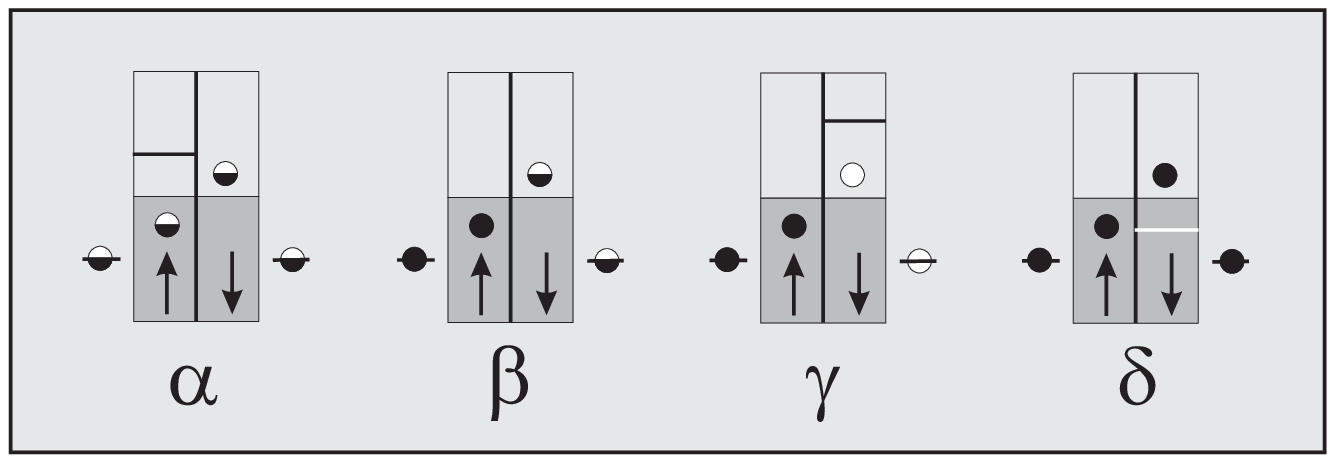}%
}%
%EndExpansion
\\
&
\begin{tabular}
[c]{l}%
Fig.2: An electron has been injected into the spin-up band and the spin up
nest.\\
This induces electron or hole transitions in the spin-down band. The
resulting\\
Slater states are shown as ($\gamma$) and ($\delta$). (Each band with half
circles consists\\
of two Slater states.)
\end{tabular}
\end{align*}

The injection into $a_{0\uparrow}^{\dag}$ or $d_{\uparrow}^{\dag}$ yields%
\begin{equation}%
\begin{tabular}
[c]{lll}%
$a_{0\uparrow}^{\dag}\Psi_{MS}=\left(  A_{d,b}a_{0}^{\dag}d_{\uparrow}^{\dag
}b_{0\downarrow}^{\dag}+A_{d,d}a_{0}^{\dag}d_{\uparrow}^{\dag}d_{\downarrow
}^{\dag}\right)  \left\vert \mathbf{0}_{a,\uparrow}\mathbf{0}_{b,\downarrow
}\right\rangle $ & for & $a_{0}^{\dag}$\\
$d_{\uparrow}^{\dag}\Psi_{MS}=-\left(  A_{a,b}a_{0\uparrow}^{\dag}d_{\uparrow
}^{\dag}b_{0\downarrow}^{\dag}+A_{a,d}a_{0\uparrow}^{\dag}d_{\uparrow}^{\dag
}d_{\downarrow}^{\dag}\right)  \left\vert \mathbf{0}_{a,\uparrow}%
\mathbf{0}_{b,\downarrow}\right\rangle $ & for & $d_{\uparrow}^{\dag}$%
\end{tabular}
\ \ \ \ \label{a_Psi_MS}%
\end{equation}
Both final states yield a double occupancy of the spin up nest states.
Furthermore these states are not eigenstates of the nest. With respect to the
(basis) states $a_{0}^{\dag}d_{\uparrow}^{\dag}b_{0\downarrow}^{\dag
}\left\vert \mathbf{0}_{a,\uparrow}\mathbf{0}_{b,\downarrow}\right\rangle $
and $a_{0}^{\dag}d_{\uparrow}^{\dag}d_{\downarrow}^{\dag}\left\vert
\mathbf{0}_{a,\uparrow}\mathbf{0}_{b,\downarrow}\right\rangle $ the nest
Hamiltonian takes the form%
\begin{equation}
H_{2/1}^{nst}=\left(
\begin{array}
[c]{cc}%
\left(  E_{d}+E_{0}^{\left(  a\right)  }\right)  +E_{0}^{\left(  b\right)  } &
V_{b}^{sd}\\
V_{b}^{sd} & \left(  E_{d}+E_{0}^{\left(  a\right)  }\right)  +E_{d}+U
\end{array}
\right)  \label{Hnst_3}%
\end{equation}
By diagonalization one obtains the eigenstates $\Psi_{2/1}^{\left(  1\right)
}$ and $\Psi_{2/1}^{\left(  2\right)  }$ with
\begin{equation}
\Psi_{2/1}^{\left(  \alpha\right)  }=a_{0\uparrow}^{\dag}d_{\uparrow}^{\dag
}\left(  B_{b\downarrow}^{\left(  \alpha\right)  }b_{0\downarrow}^{\dag
}+B_{d\downarrow}^{\left(  \alpha\right)  }d_{\downarrow}^{\dag}\right)
\left\vert \mathbf{0}_{a,\uparrow}\mathbf{0}_{b,\downarrow}\right\rangle
\label{Psi2/1}%
\end{equation}

The states ($\alpha$) and ($\beta$) in Fig.2 are the initial states which one
obtains through injection of a spin-up electron into the ground state. Due to
the perturbation Hamiltonian $H_{1}^{\prime}$ these states interact with each
other and ($\beta$) interacts with the states ($\gamma$) and ($\delta$).

In table I the possible states which can be obtained through the injection of
a spin-up electron plus linear coupling through $H_{1}^{\prime}$ are
collected. These states are ($\alpha$) the two-electron nest ground state plus
one electron, ($\beta$) a nest with two spin-up and one spin-down electron,
($\gamma$) a full spin-up and empty spin-down nest plus one spin down electron
and ($\delta$) a full spin-up and full spin-down nest and one spin down hole.
In table I these states, their number and their energies are listed.%

\begin{align*}
&
\begin{tabular}
[c]{|l|l|l|l|}\hline
$\mathbf{\Psi}_{f}$ & \textbf{number} & \textbf{energy} & \\\hline
$a_{j\uparrow}^{\dag}\Psi_{MS}$ & $N/2$ & $E_{j}^{\left(  a\right)  }$ &
\\\hline
$\Psi_{2/1}^{\left(  \alpha\right)  }$ & $2,\alpha=1,2$ & $E_{d}%
+E_{0}^{\left(  a\right)  }+E_{2/1}^{\alpha}-E_{00}$ & \\\hline
$a_{0\uparrow}^{\dag}d_{\uparrow}^{\dag}b_{j\downarrow}^{\dag}\left\vert
\mathbf{0}_{a,\uparrow}\mathbf{0}_{b,\downarrow}\right\rangle $ & $N/2$ &
$Ed+E_{0}^{\left(  a\right)  }+E_{j}^{\left(  b\right)  }-E_{00}$ & \\\hline
$b_{k\downarrow}a_{0\uparrow}^{\dag}d_{\uparrow}^{\dag}b_{0\downarrow}^{\dag
}d_{\downarrow}^{\dag}\left\vert \mathbf{0}_{a,\uparrow}\mathbf{0}%
_{b,\downarrow}\right\rangle $ & $N/2-1$ & $%
\begin{array}
[c]{c}%
2E_{d}+U+E_{0}^{\left(  a\right)  }+E_{0}^{\left(  b\right)  }\\
-E_{k}^{\left(  b\right)  }-E_{00}%
\end{array}
$ & \\\hline
\end{tabular}
\\
&
\begin{tabular}
[c]{l}%
Table I: This table describes the states of Fig.2 which are generated\\
by the injection of one spin-up electron into the magnetic ground state\\
and transition from the resulting states through $H_{1}^{\prime}$. The state
$\Psi_{MS}$ is\\
given by equ. (\ref{Psi_MS}) and $\Psi_{2/1}^{\left(  \alpha\right)  }$ is
given by (\ref{Psi2/1}). The\\
energy is measured from the ground-state energy $E_{00}$.
\end{tabular}
\end{align*}

Fig.2 and table I show all the spin-up electron excitations which interact
linearly in $H_{1}^{\prime}$. This is a total of $\left(  \frac{3}%
{2}N+1\right)  $ states. It is straight forward to construct the secular
matrix (i.e. the excitation Hamiltonian $H^{xct}$) between the excitations in
table I. One may put the two nest states $\Psi_{2/1}^{\left(  1\right)  }$ and
$\Psi_{2/1}^{\left(  2\right)  }$ at the positions one and two, followed by
the $\left(  \frac{3}{2}N-1\right)  $ additional single particle excitations.
We denote these $\left(  \frac{3}{2}N+1\right)  $ states as $\varphi_{\nu}$.
The diagonal of the Hamiltonian is given by the energies in table I. The
off-diagonal elements of $H^{xct}$ are the matrix elements of $H_{1}^{\prime}$
between the states ($\alpha,\beta,\gamma,\delta$) in Fig.2. The single
particle excitations interact only with the first two nest states through
$H_{1}^{\prime}$ but not among each other. (In appendix B the corresponding
matrix elements are shown in table III for similar transitions from the ground state.)

This Hamiltonian $H^{xct}$ is diagonalized and yields a set of $\left(
\frac{3}{2}N+1\right)  $ new eigenstates $\psi_{\mu}$ with eigenenergies
$E_{\mu}^{xct}$. The components of the eigenstates $\psi_{\mu}$ in terms of
$\varphi_{\nu}$ are given as columns $\left(  \psi_{\mu}^{\nu}\right)  $. The
$\nu$th row of the matrix $\left(  \psi_{\mu}^{\nu}\right)  $ yields the
amplitude of our $\nu$th original state $\varphi_{\nu}$ in terms of the new
eigenstates $\psi_{\mu}$. If this $\nu$th original state is, for example,
$a_{j\uparrow}^{\dag}\Psi_{MS}$ then it can be expressed in the new
eigenstates as
\[
\varphi_{\nu}=a_{j\uparrow}^{\dag}\Psi_{MS}=%
%TCIMACRO{\tsum _{\mu}}%
%BeginExpansion
{\textstyle\sum_{\mu}}
%EndExpansion
\psi_{\mu}^{\nu}\psi_{\mu}%
\]
Its density of states is then
\[
N_{\nu}\left(  \varepsilon\right)  =%
%TCIMACRO{\tsum _{\mu}}%
%BeginExpansion
{\textstyle\sum_{\mu}}
%EndExpansion
\left\vert \psi_{\mu}^{\nu}\right\vert ^{2}\delta\left(  \varepsilon-E_{\mu
}^{xct}\right)
\]

Since electron injection creates only the states ($\alpha$) and ($\beta$) one
obtains the full (spin-up) excitation spectrum by summing over these $\left(
\frac{1}{2}N+2\right)  $ states. The weight of states $a_{j\uparrow}^{\dag
}\Psi_{MS}$ is one, however, the weight of $a_{0\uparrow}^{\dag}\Psi_{MS}$ is
only $\left\vert A_{d,b}\right\vert ^{2}+\left\vert A_{d,d}\right\vert ^{2}$
since
\begin{align*}
&  a_{0\uparrow}^{\dag}\left[  A_{a,b}a_{0\uparrow}^{\dag}b_{0\downarrow
}^{\dag}+A_{a,d}a_{0\uparrow}^{\dag}d_{\downarrow}^{\dag}+A_{d,b}d_{\uparrow
}^{\dag}b_{0\downarrow}^{\dag}+A_{d,d}d_{\uparrow}^{\dag}d_{\downarrow}^{\dag
}\right]  \left\vert \mathbf{0}_{a,\uparrow}\mathbf{0}_{b,\downarrow
}\right\rangle \\
&  =\left[  A_{d,b}a_{0\uparrow}^{\dag}d_{\uparrow}^{\dag}b_{0\downarrow
}^{\dag}+A_{d,d}a_{0\uparrow}^{\dag}d_{\uparrow}^{\dag}d_{\downarrow}^{\dag
}\right]  \left\vert \mathbf{0}_{a,\uparrow}\mathbf{0}_{b,\downarrow
}\right\rangle
\end{align*}
A similar result is found for the weight of $d_{\uparrow}^{\dag}\Psi_{MS}$
which is $\left\vert A_{a,b}\right\vert ^{2}+\left\vert A_{a,d}\right\vert
^{2}$.

Since $a_{0\uparrow}^{\dag}\Psi_{MS}$ and $d_{\uparrow}^{\dag}\Psi_{MS}$ are
not eigenstatet of $H_{0}^{\prime}$ they represent a combination of the two
eigenstates $\Psi_{2/1}^{\left(  \alpha\right)  }$. In Fig.3 is sketched what
happens when an electron is injected into either the state $a_{0\uparrow
}^{\dag}$ or $d_{\uparrow}^{\dag}$. The electron injection yields a
superposition of the two eigenstates $\Psi_{2/1}^{\left(  \alpha\right)  }$.
From these states the electron can make a transition into any of the ($\gamma
$) states via $H_{1}^{\prime}$. The two transition amplitudes interfere in
this transition. This interference has to be included in the calculation of
the spectral weight density of $a_{0\uparrow}^{\dag}\Psi_{MS}$ and
$d_{\uparrow}^{\dag}\Psi_{MS}$ (which requires just the scalar products
between $a_{0\uparrow}^{\dag}\Psi_{MS}$ and $\Psi_{2/1}^{\left(
\alpha\right)  }$ (or $d_{\uparrow}^{\dag}\Psi_{MS}$ and $\Psi_{2/1}^{\left(
\alpha\right)  }$). This is discussed in more detail in appendix C.%

\begin{align*}
&
%TCIMACRO{\FRAME{itbpF}{5.056in}{2.2648in}{0in}{}{}{corl181_{2}g.eps}%
%width 5.056in;  height 2.2648in;  depth 0in;  original-width 6.5056in;
%original-height 2.8941in;  cropleft "0";  croptop "1";  cropright "1";
%cropbottom "0";  filename 'Fig3.eps';file-properties "XNPEU";}}}%
%BeginExpansion
{\includegraphics[
height=2.2648in,
width=5.056in
]%
{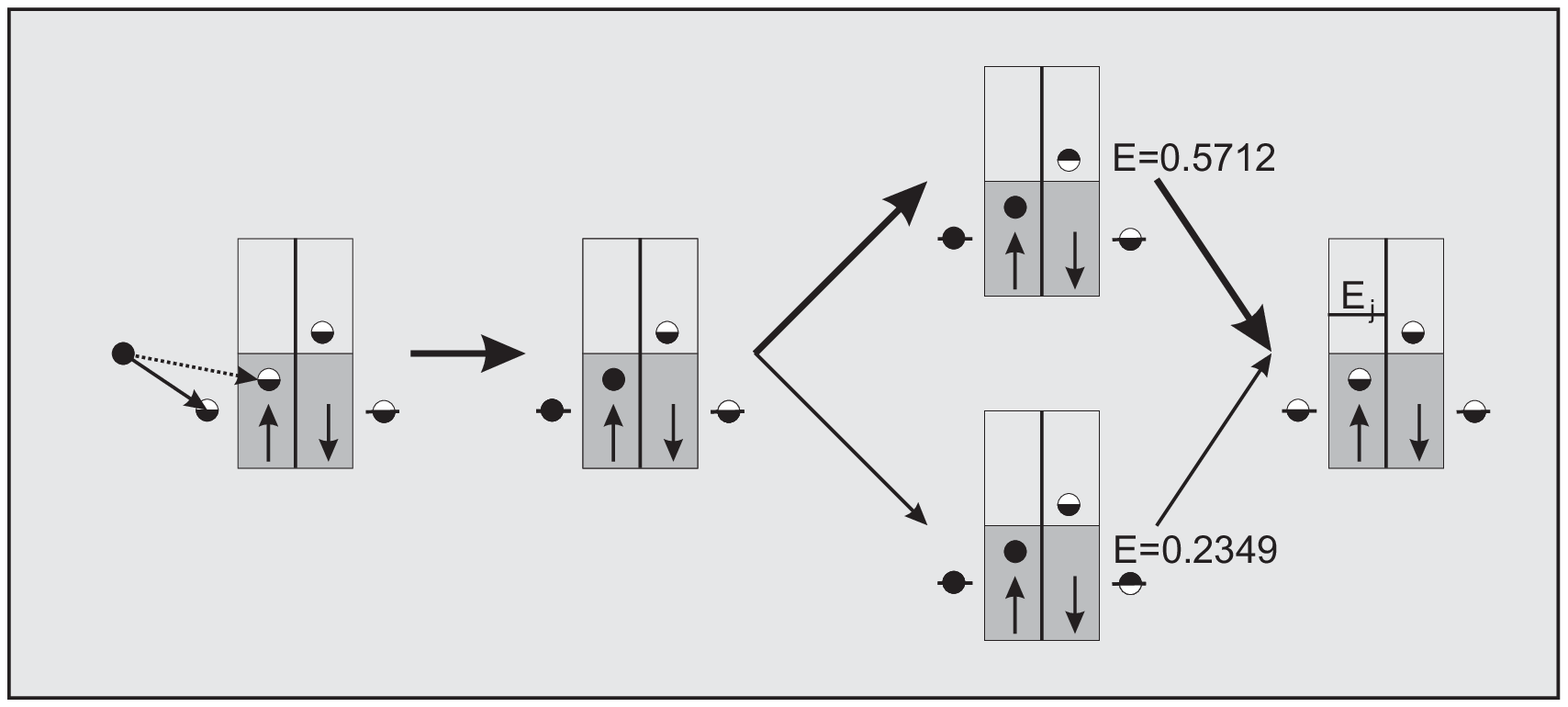}%
}%
%EndExpansion
\\
&
\begin{tabular}
[c]{l}%
Fig.3: An electron has been injected into the $a_{0\uparrow}^{\dag}$ state (or
$d_{\uparrow}^{\dag}$ state,\\
dashed arrow). The resulting state is a superposition of two nest states.\\
From these nest states the electron makes (as one possibility) a transition\\
into the state $a_{j\uparrow}^{\dag}$ where the two amplitudes interfere. The
energies of the\\
two nest states are shown. The different thickness of the arrows shows\\
different probabilities for the two paths.
\end{tabular}
\end{align*}

\subsection{Injection of a hole}

For the full spectrum of excitations one has to include the injection of holes
into the occupied states. This is shown in Fig.4. The hole can be injected
into the occupied states $a_{j\uparrow}^{\dag}$ yielding $a_{j\uparrow}%
\Psi_{MS}$ or into the nest. In the latter case the spin-up part of the nest
is emptied. This yields for the secular matrix of the nest in analogy to equ.
(\ref{Hnst_3})%
\begin{equation}
H_{0/1}^{nst}=\left(
\begin{array}
[c]{cc}%
E_{0}^{\left(  b\right)  } & V_{b}^{sd}\\
V_{b}^{sd} & E_{d}%
\end{array}
\right)  \label{Hnst_1}%
\end{equation}%
\begin{align*}
&
%TCIMACRO{\FRAME{itbpF}{5.335in}{1.8049in}{0in}{}{}{corl181_{2}e.eps}%
%width 5.335in;  height 1.8049in;  depth 0in;  original-width 5.0668in;
%original-height 1.6961in;  cropleft "0";  croptop "1";  cropright "1";
%cropbottom "0";  filename 'Fig4.eps';file-properties "XNPEU";}}}%
%BeginExpansion
{\includegraphics[
height=1.8049in,
width=5.335in
]%
{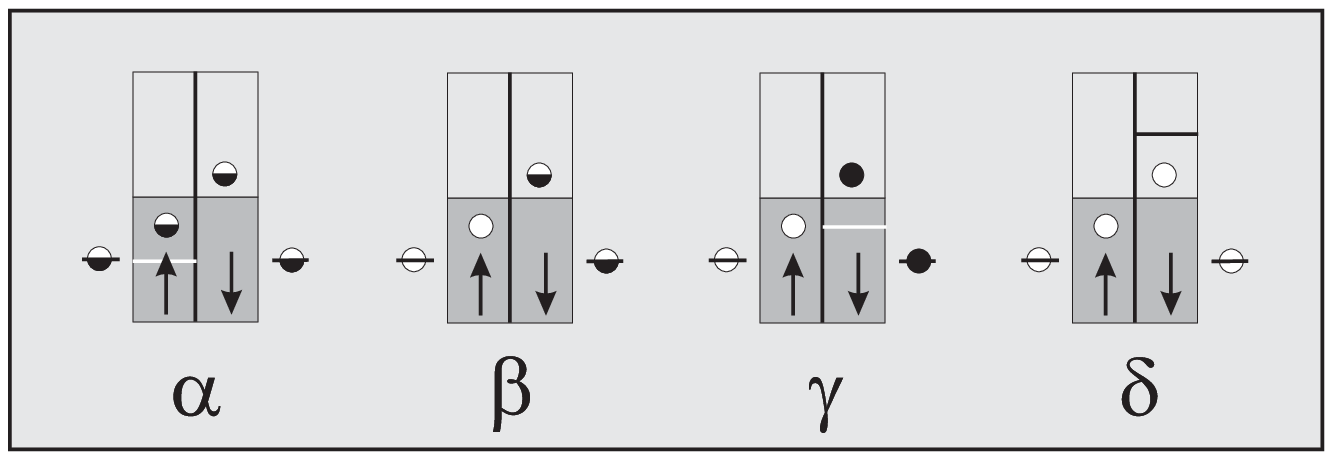}%
}%
%EndExpansion
\\
&
\begin{tabular}
[c]{l}%
Fig.4: An hole has been injected into the spin-up band and the spin up nest.\\
This induces electron or hole transitions in the spin-down band. The
resulting\\
Slater states are shown as ($\gamma$) and ($\delta$).
\end{tabular}
\end{align*}

\begin{align*}
&
\begin{tabular}
[c]{|l|l|l|}\hline
\textbf{state} & \textbf{number} & \textbf{energy}\\\hline
$a_{j\uparrow}\Psi_{MS}$ & $N/2-1$ & $-E_{j}^{\left(  a\right)  }=\left\vert
E_{j}^{\left(  a\right)  }\right\vert $\\\hline
$\Psi_{\downarrow}^{\left(  \alpha\right)  }=\left(  B_{b}b_{0\downarrow
}^{\dag}+B_{d}d_{\downarrow}^{\dag}\right)  \left\vert \mathbf{0}_{a,\uparrow
}\mathbf{0}_{b,\downarrow}\right\rangle $ & $2,\alpha=1,2$ & $E_{\downarrow
}^{\left(  \alpha\right)  }-E_{00}$\\\hline
$b_{j\downarrow}b_{0\downarrow}^{\dag}d_{\downarrow}^{\dag}\left\vert
\mathbf{0}_{a,\uparrow}\mathbf{0}_{b,\downarrow}\right\rangle $ & $N/2-1$ &
$E_{d}+E_{0}^{\left(  b\right)  }-E_{j}^{\left(  b\right)  }-E_{00}$\\\hline
$b_{j\downarrow}^{\dag}\left\vert \mathbf{0}_{a,\uparrow}\mathbf{0}%
_{b,\downarrow}\right\rangle $ & $N/2$ & $E_{j}^{\left(  b\right)  }-E_{00}%
$\\\hline
\end{tabular}
\\
&
\begin{tabular}
[c]{l}%
Table II: This table describes the states of Fig.4 which are generated\\
by the injection of one spin-up hole into the magnetic ground state\\
and transition from the resulting states through $H_{1}^{\prime}$. The energy
is\\
measured from the ground-state energy $E_{00}$.
\end{tabular}
\end{align*}

The construction of the excitation or secular Hamiltonian $H^{xct}$ is in
complete analogy to the electron injection. This time the number of
excitations is $\frac{3}{2}N$. The spectrum is obtained in the same way as before.

\section{Numerical Results}

For the numerical calculation a conduction band with a finite number of states
is used. We follow here Wilson \cite{W18} by using an s-electron band with
constant density of states and the Fermi level in the center, which we divide
into energy cells $\mathfrak{C}_{\nu}$. In each energy cell (which may contain
$Z_{\nu}$ $\mathbf{k}$-states $c_{\mathbf{k}}^{\dag})$ we rearrange the states
(by an orthogonal transformation) so that one state $c_{\nu}^{\dag}=Z_{\nu
}^{-1/2}%
%TCIMACRO{\tsum _{\mathfrak{C}_{\nu}}}%
%BeginExpansion
{\textstyle\sum_{\mathfrak{C}_{\nu}}}
%EndExpansion
c_{\mathbf{k}}^{\dag}$ accumulates all the interaction with the d-states while
the other $\left(  Z_{\nu}-1\right)  $ states have zero interaction with the
d-states. Wilson normalized the energy in terms of the Fermi energy so that
his band extended from $-1$ $\ $to $1$. Wilson's logarithmic scale for the
energy cells is not opportune for the present investigation because it is not
fine enough at the energy of the d-resonance. Therefore I use a linear
sub-division of the energy band $\left(  -1:1\right)  $. For the majority of
calculations the energy band is sub-divided into $N=40,80$ and $160$ energy
cells. The state $c_{\nu}^{\dag}$ represents all the s-electron states in the
cell $\mathfrak{C}_{\nu}=\left(  -1+\nu\frac{2}{N}:-1+\left(  \nu+1\right)
\frac{2}{N}\right)  $ and possesses the average energy $\varepsilon_{\nu
}=-1+\left(  \nu+\frac{1}{2}\right)  \frac{2}{N}$ (corresponding to
$-.975,-.925,..+.975$ for $N=40$). For $N$ energy cells with constant width of
$2/N$ the s-d-matrix elements $V_{\nu}^{sd}$ is given by $V_{sd}^{0}/\sqrt{N}$
so that $%
%TCIMACRO{\tsum _{\nu}}%
%BeginExpansion
{\textstyle\sum_{\nu}}
%EndExpansion
\left\vert V_{\nu}^{sd}\right\vert ^{2}=\left\vert V_{sd}^{0}\right\vert ^{2}$.

In the following I show the results for the Friedel-Anderson Hamiltonian with
the d-level energy $E_{d}=-0.5,$ the Coulomb energy $U=1.0$ and an s-d-hopping
matrix element of $\left\vert V_{sd}^{0}\right\vert ^{2}=0.025$. The magnetic
moment of this impurity is $\mu=0.998\mu_{B}$ (for $N=80$). The calculations
are performed with $N=40,80$ or $160$ energy levels of constant spacing. A
second set of results is derived for the parameters $E_{d}=-0.5$, $U=1$ and
$\left\vert V_{sd}^{0}\right\vert ^{2}=0.05$.

To assure that the magnetic state $\Psi_{MS}$ is indeed the ground state (i.e.
to prevent the formation of a singlet state) a magnetic field $B$ is applied
yielding a magnetic energy $E_{B}=\mu_{B}B$. This energy is chosen so that
$E_{B}>10k_{B}T_{K}$. I estimate the Kondo energy from the difference between
the energies of the singlet and the triplet state. This energy difference is
about $8\times10^{-4}$ for $\left\vert V_{sd}^{0}\right\vert ^{2}=0.05$ and of
the order of $10^{-7}$ for $\left\vert V_{sd}^{0}\right\vert ^{2}=0.025$. It
turned out that the required magnetic energy is in both cases so small that it
yields no noticeable changes. This is partly due to the fact that the absolute
smallest band energies which are $\pm1/N$ act as a finite temperature as
Wilson pointed out \cite{W18}. For $N=40$ this corresponds to a temperature of
$\varepsilon_{F}/40$ which is a very large temperature compared with most
Kondo temperatures. But magnetic field is of academic importance to assure the
magnetic state $\Psi_{MS}$ is the appropriate ground state.

The use of equidistant energy levels is important to identify the resonance
state within the electron bands. But it has the drawback that it does not
describe well the behavior of the wave function at low energies. At low
energies the logarithmic energy scale which Wilson introduced would be more
appropriate. But the evaluation of the density of states is much more
difficult for a non-linear energy scale.

For each spin band one obtains a spectrum with a total weight of $\left(
N+1\right)  $, corresponding to $N$ s-electron states and one d-electron
state. However, the number of energy levels is $\left(  3N+1\right)  $. (This
is the number of eigenstates of the excitation Hamiltonians for electrons and
holes together). This means that the weight at the individual energies is at
least for $2N$ energies much less than one. In Fig.5 the spectral weight at
different energies is shown in the energy range from $-1$ to $+1$ for the
minority band. For negative energies the weight is either very close to one or
very small. Here one can calculate the density of states by the separation of
the levels with weight close to one (by dividing the weight by the level
separation). The evaluation is more complicated for positive energies.
However, here the sum of neighboring energy levels is close to one. Then one
can calculated the "center of weight" for two neighboring levels which have a
total weight close to one and then proceed as before. It turns out that the
best approach is to start from the lower and upper ends of the band in the evaluation.%

\begin{align*}
&
%TCIMACRO{\FRAME{itbpF}{4.0598in}{3.2179in}{0in}{}{}{org181_{5}d.eps}%
%width 4.0598in;  height 3.2179in;  depth 0in;  original-width 3.8497in;
%original-height 3.0461in;  cropleft "0";  croptop "1";  cropright "1";
%cropbottom "0";  filename 'Fig5.eps';file-properties "XNPEU";}}}%
%BeginExpansion
{\includegraphics[
height=3.2179in,
width=4.0598in
]%
{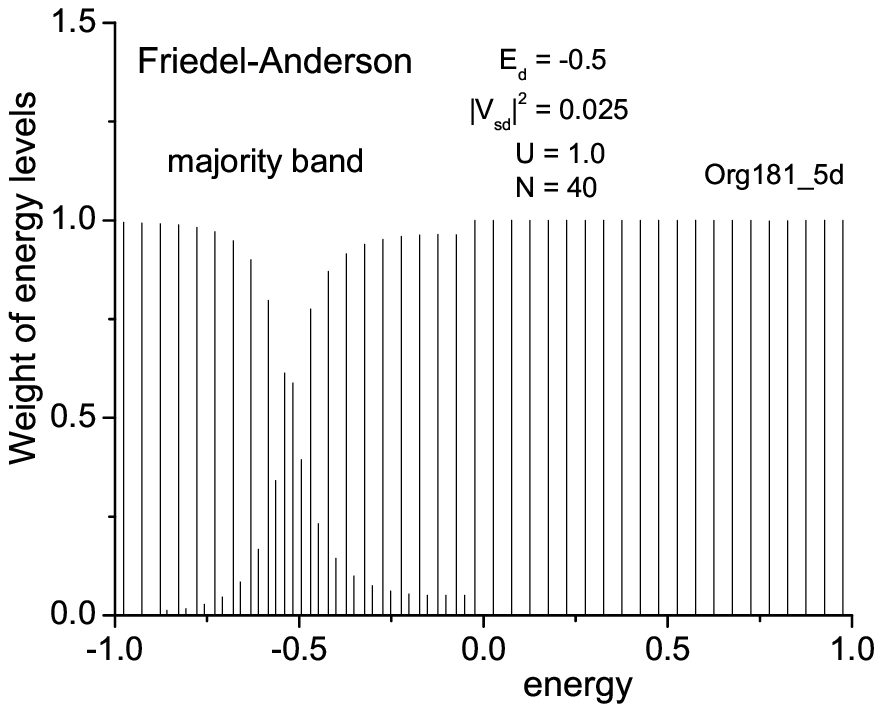}%
}%
%EndExpansion
\\
&
\begin{tabular}
[c]{l}%
Fig.5: The spectral weight for the different excitation energies.
\end{tabular}
\end{align*}

In Fig.6a the density of states of the excitation spectrum for the majority
spin is shown. The full circles are obtained with $N=40$ states and the stars
use $N=80$ equally spaced Wilson states. The full curve represents a Lorentz
curve with the parameters
\[
N_{d}\left(  \varepsilon\right)  =\frac{1}{\pi}\frac{0.08}{\left(
\varepsilon-\left(  -.53\right)  \right)  ^{2}+0.08^{2}}%
\]
The resonance energy is is $E_{r}=-0.53$ and the resonance (half) width is
$\Gamma_{r}=0.08$. The corresponding mean field resonance width is
$\Gamma_{mf}=0.039$.%

\begin{align*}
&
%TCIMACRO{\FRAME{itbpF}{4.1859in}{3.3018in}{0in}{}{}{org181_{1}1a.eps}%
%width 4.1859in;  height 3.3018in;  depth 0in;  original-width 3.9701in;
%original-height 3.1249in;  cropleft "0";  croptop "1";  cropright "1";
%cropbottom "0";  filename 'Fig6a.eps';file-properties "XNPEU";}}}%
%BeginExpansion
{\includegraphics[
height=3.3018in,
width=4.1859in
]%
{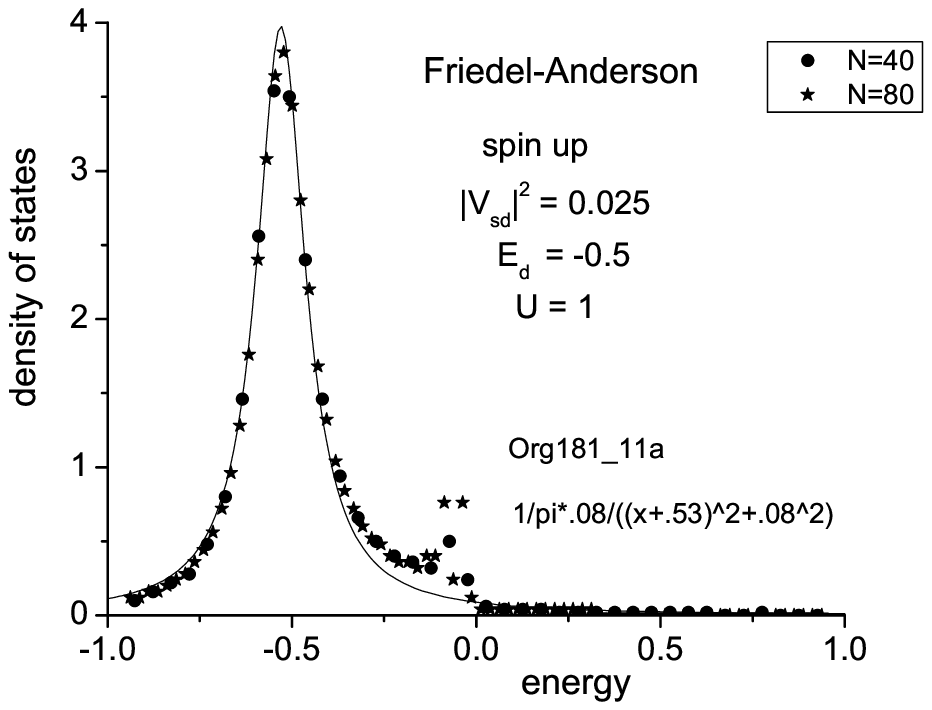}%
}%
%EndExpansion
\\
&
\begin{tabular}
[c]{l}%
Fig.6a: The density of states for the majority spins. The full circles\\
are calculated with $N=40$ states and the stars use $N=80$ equally\\
spaced Wilson states. The full curve represents a Lorentz curve\\
with the resonance energy $E_{r}=-0.53$ and the width $\Gamma_{r}=0.08$.\\
This width is twice the mean-field value of $\Gamma_{mf}=0.039$.
\end{tabular}
\end{align*}

In Fig.6b the density of states of the minority spin is drawn. Again the full
curve represents a Lorentz curve with the resonance at $E_{r}=0.52$ and a
resonance half-width of $\Gamma_{r}=0.8$.%

\begin{align*}
&
%TCIMACRO{\FRAME{itbpF}{4.0598in}{3.3018in}{0in}{}{}{org181_{1}1b.eps}%
%width 4.0598in;  height 3.3018in;  depth 0in;  original-width 3.8497in;
%original-height 3.1249in;  cropleft "0";  croptop "1";  cropright "1";
%cropbottom "0";  filename 'Fig6b.eps';file-properties "XNPEU";}}}%
%BeginExpansion
{\includegraphics[
height=3.3018in,
width=4.0598in
]%
{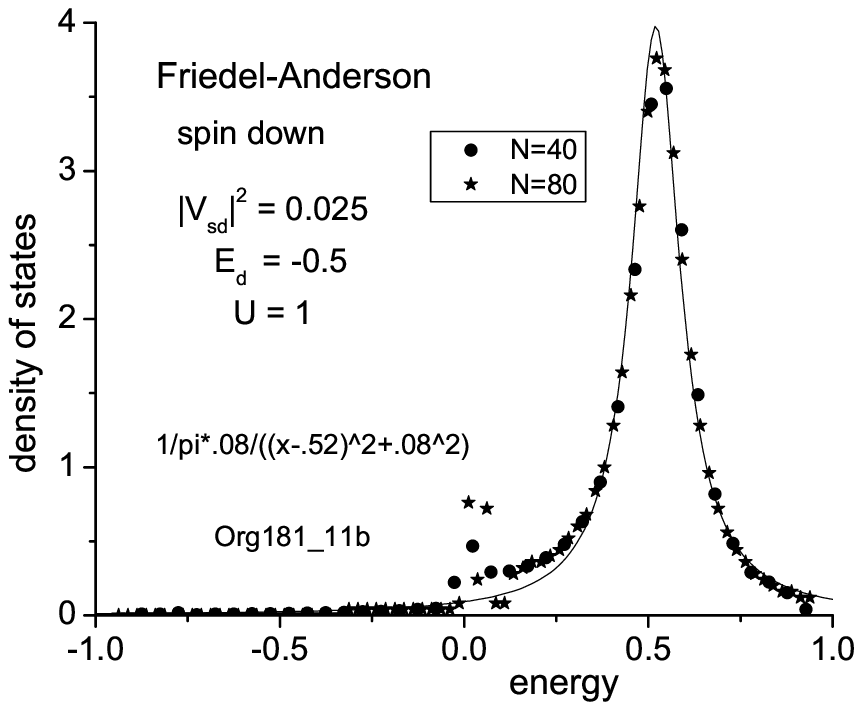}%
}%
%EndExpansion
\\
&
\begin{tabular}
[c]{l}%
Fig.6b: The density of states for the minority spins. The full circles\\
are calculated with $N=40$ states and the stars use $N=80$ equally\\
spaced Wilson states. The full curve represents a Lorentz curve\\
with the resonance energy $E_{r}=0.52$ and the width $\Gamma_{r}=0.08$.\\
Again this width is twice the mean-field value of $\Gamma_{mf}=0.039$.
\end{tabular}
\end{align*}

For comparison Fig.7 shows the result of a similar calculation and evaluation
for a simple Friedel resonance, where the Coulomb energy $U$ is set equal to
zero. The evaluation yields a Lorentz curve with $E_{r}=-0.51$ and $\Gamma
_{r}=0.044.$ The s-d-matrix element is still given by $\left\vert V_{sd}%
^{0}\right\vert ^{2}=0.025$.%

\begin{align*}
&
%TCIMACRO{\FRAME{itbpF}{3.9493in}{3.3018in}{0in}{}{}{org181_{1}1c.eps}%
%width 3.9493in;  height 3.3018in;  depth 0in;  original-width 3.7443in;
%original-height 3.1249in;  cropleft "0";  croptop "1";  cropright "1";
%cropbottom "0";  filename 'Fig7.eps';file-properties "XNPEU";}}}%
%BeginExpansion
{\includegraphics[
height=3.3018in,
width=3.9493in
]%
{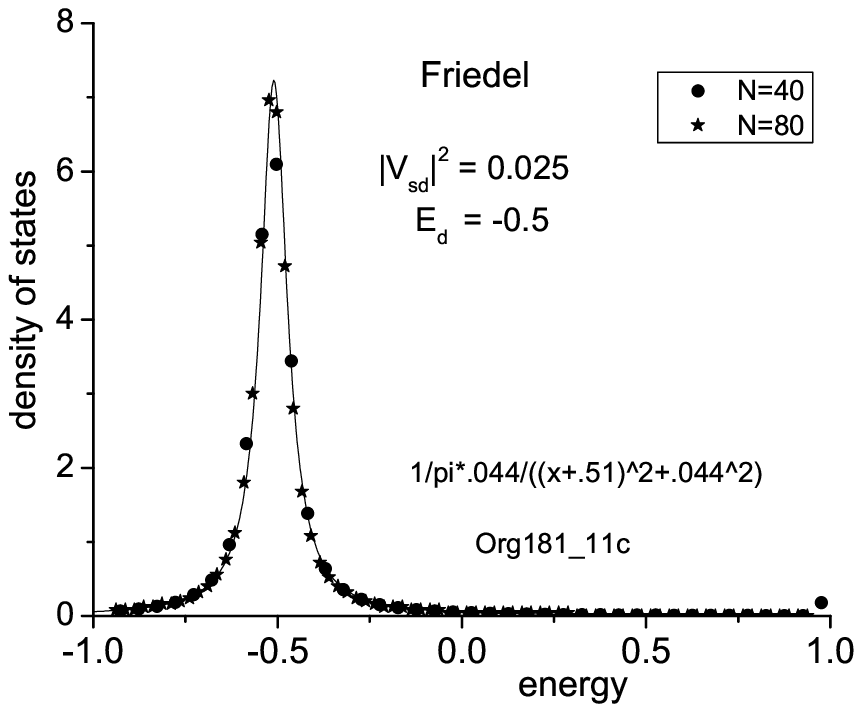}%
}%
%EndExpansion
\\
&
\begin{tabular}
[c]{l}%
Fig.7: The density of states for an impurity with $U=0$. The full circles\\
and the stars are calculated with $N=40$ and $N=80$ states.\\
The full curve represents a Lorentz curve with the resonance energy\\
$E_{r}=-0.51$ and the width $\Gamma_{r}=0.044$.
\end{tabular}
\end{align*}

Alternatively we tried to obtain the density of states by broadening the
$\delta$-shaped energy spectrum in Fig.5 with a Gaussian curve $\sqrt{1/2\pi
}\exp\left[  \left(  \varepsilon-E_{d}\right)  ^{2}/2\sigma^{2}\right]  $.
This method worked quite well. The optimal density curve was obtained when
$\sqrt{2}\sigma$ was equal to the level distance $2/N$. The width of the two
resonances was essentially the same as in Fig.6a,b and Fig.8a,b. Only the
heights were slightly reduced. However, I prefer to use the other evaluation
method in this paper so that there is no doubt that the broadening of the
resonance width is a real physical effect and not due to an artificial
broadening with a Gaussian curve.

In a second series the majority and minority density of states have been
calculated for the parameters $E_{d}=-0.5$, $U=1$, and $\left\vert V_{sd}%
^{0}\right\vert ^{2}=0.05$. This impurity possesses a magnetic moment of
$\mu=0.997\mu_{B}$. In this case the mean-field theory yields a resonance
width of $\Gamma_{mf}$=0.0$7\allowbreak9$. The best fit to the numerical
results yields $\Gamma_{FAIR}=0.17$. It is again about twice the value of the
mean field. (The resonance curves are no longer perfectly symmetrical because
of the finite width of the conduction band).

The corresponding Friedel density of states (which is not shown here) has a
resonance with of 0.08 which is quite close to the theoretical value of
$0.079$.
\begin{align*}
&
%TCIMACRO{\FRAME{itbpF}{4.2565in}{3.3018in}{0in}{}{}{org181_{1}2a.eps}%
%width 4.2565in;  height 3.3018in;  depth 0in;  original-width 4.0373in;
%original-height 3.1249in;  cropleft "0";  croptop "1";  cropright "1";
%cropbottom "0";  filename 'Fig8a.eps';file-properties "XNPEU";}}}%
%BeginExpansion
{\includegraphics[
height=3.3018in,
width=4.2565in
]%
{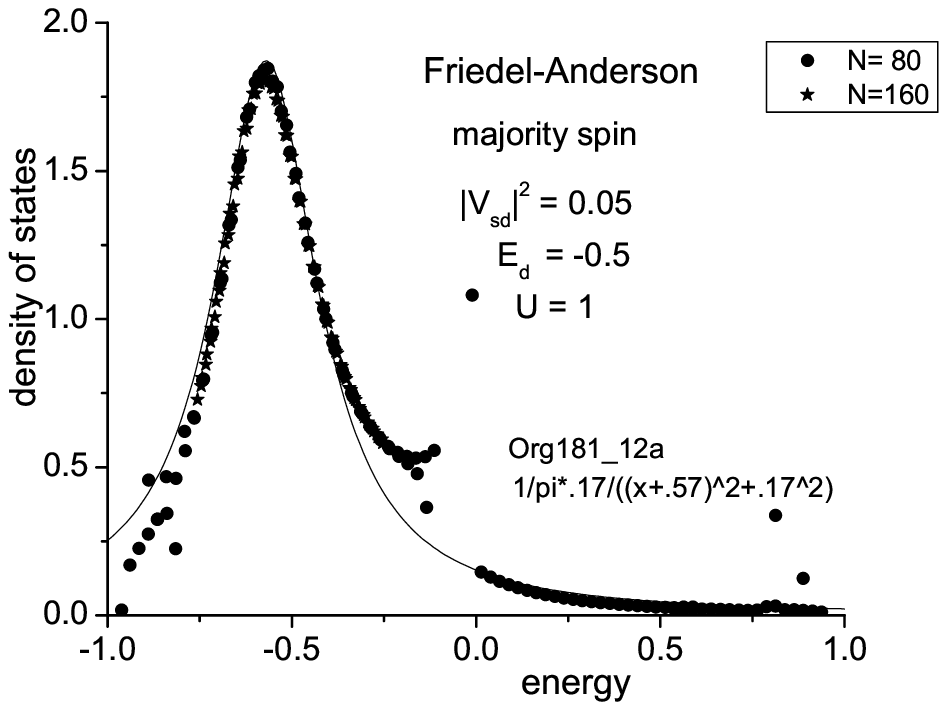}%
}%
%EndExpansion
\\
&
\begin{tabular}
[c]{l}%
Fig.8a: The density of states for the majority spins for $\left\vert
V_{sd}^{0}\right\vert ^{2}=0.05$%
\end{tabular}
\end{align*}%
\begin{align*}
&
%TCIMACRO{\FRAME{itbpF}{4.0598in}{3.3018in}{0in}{}{}{org181_{1}2b.eps}%
%width 4.0598in;  height 3.3018in;  depth 0in;  original-width 3.8497in;
%original-height 3.1249in;  cropleft "0";  croptop "1";  cropright "1";
%cropbottom "0";  filename 'Fig8b.eps';file-properties "XNPEU";}}}%
%BeginExpansion
{\includegraphics[
height=3.3018in,
width=4.0598in
]%
{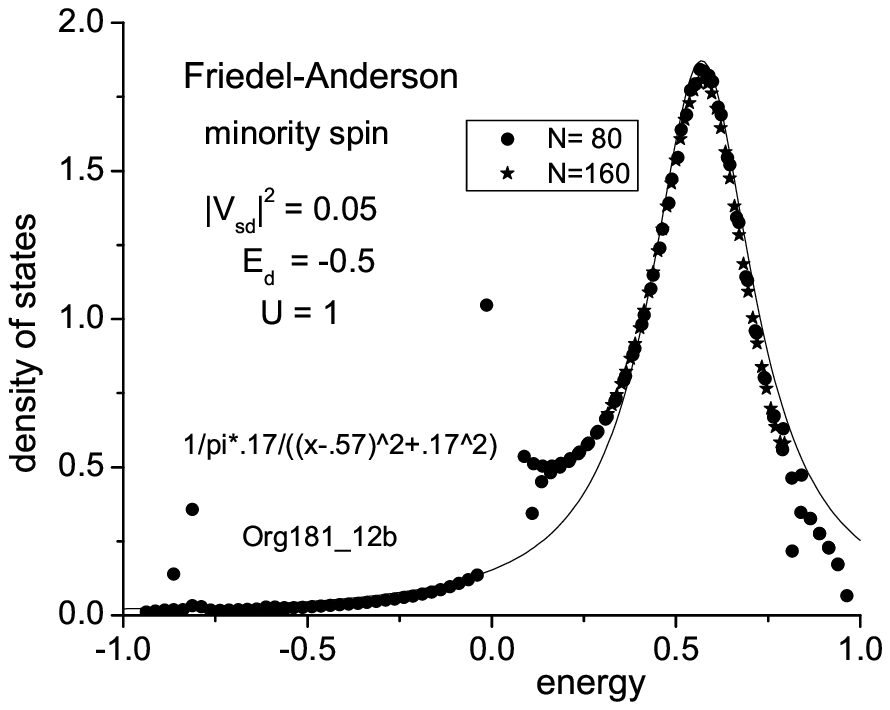}%
}%
%EndExpansion
\\
&
\begin{tabular}
[c]{l}%
Fig.8b: The density of states for the minority spins for $\left\vert
V_{sd}^{0}\right\vert ^{2}=0.05$%
\end{tabular}
\end{align*}%
\[
\]

\section{Discussion}

In the mean field approximation the magnetic state has two d-resonances at the
energies $E_{d,\sigma}=E_{d}+Un_{d,-\sigma}$. Since in the symmetric case one
has $n_{d,\sigma}=\left(  1\mp\mu\right)  /2$ one finds $E_{d,\sigma}%
=E_{d}+U\left(  1\mp\mu\right)  /2=\mp U\mu/2=\pm\mu E_{d}$ since
$E_{d}+U/2=0$. So generally the resonances are closer to Fermi energy than
$\pm E_{d}.$ In our case for the parameters $E_{d}=-0.5,$ $U=1$ and
$\left\vert V_{sd}^{0}\right\vert ^{2}=0.025$ the magnetic moment is within
1\% $\mu\thickapprox1$ (in units of $\mu_{B})$ and one expects the resonance
almost at $\pm0.5$. The resonance width in mean field is given by $\Gamma
_{mf}=\pi\left\vert V_{sd}^{0}\right\vert ^{2}g\left(  \varepsilon_{F}\right)
$ where $g\left(  \varepsilon_{F}\right)  $ is the density of states of the
s-electrons at the Fermi level. This yields for the above parameters
$\Gamma_{mf}=\pi\ast0.025\ast.5=\allowbreak0.03\allowbreak9$. (The matrix
element $V_{sd}^{0}$ and the density of states are normalized to the atomic
volume as the sample volume). This is the same width that one expects for a
Friedel resonance with $\left\vert V_{sd}^{0}\right\vert ^{2}=0.025$. Indeed
in Fig.7 the Friedel resonance has a $\Gamma$-value of $\Gamma_{F}=0.044$.
This agrees within 10\% or $0.005$ with the numerical result.

The important result is that the resonance width of the minority and majority
spins is larger than the Friedel resonance width by a factor of two. Therefore
the obtained resonance width is also twice the mean field resonance width.
This suggests that the any calculation which uses mean field yields an
incorrect density of states. It will be interesting to check how the
spin-density functional theory is affected by this result because the latter
uses the mean field approximation.

Since the Coulomb interaction broadens the d-resonance by a factor of two it
also reduces the height of the resonance by the same factor of two. Therefore
it is very plausible that the mean field theory overestimates the tendency to
form a magnetic moment. If the criterion for the formation of a magnetic
moment, $UN_{d}>1$, is accepted then one expects that the critical Coulomb
energy for the formation of a magnetic moment is increased by a factor of two.
This was the previous result by the author \cite{B152}.

In the density of states of the majority and minority spins in Fig.6ab and
Fig.8a,b one observes a scattering and a small maximum at zero energy. This is
probably due to the fact that I used a constant cell width for the Wilson
states. This means that I average over all states within an energy cell of the
width $\delta E=2/N$. This is definitely a poor approximation for the two
energy cells $\mathfrak{C}_{N/2-1}$ and $\mathfrak{C}_{N/2}$ (which touch the
Fermi level). Wilson avoids this problem by using a logarithmic energy scale.
However, the present method to evaluate the density of states does not work
for an energy-dependent cell width. Details of this question will have to be
clarified in the future.

\section{Conclusion}

In this paper the density of states of the Friedel-Anderson impurity is
calculated in the magnetic ground state. The magnetic ground state is enforced
by the application of a magnetic field whose Zeeman energy is an order of
magnitude larger than the Kondo energy. (For the parameters chosen in the
numerical calculation the effect of the magnetic field is so small that it can
be neglected). The FAIR ground state is the eigenstate of a Hamiltonian
$H_{0}^{\prime}$. To construct this Hamiltonian two artificial Friedel
resonance states $a_{0,\uparrow}^{\dag}$ and $b_{0,\downarrow}^{\dag}$ are
reverse engineered out of the original spin-up and spin-down conduction bands.
There is a numerical procedure to optimize these two FAIR states. When this is
done one has an extremely simple Hamiltonian $H_{0}^{\prime}$. In
$H_{0}^{\prime}$ only the two s-states $a_{0,\uparrow}^{\dag}$ and
$b_{0,\downarrow}^{\dag}$ interact with the d-state. These states
$a_{0,\uparrow}^{\dag},$ $b_{0,\downarrow}^{\dag}$, $d_{\uparrow}^{\dag}$ and
$d_{\downarrow}^{\dag}$ are called the nest states. $H_{0}^{\prime}$ is
diagonal in the (modified) conduction band electrons $\left\{  a_{j,\uparrow
}^{\dag}\right\}  $ and $\left\{  b_{j,\downarrow}^{\dag}\right\}  $.

A perturbation Hamiltonian $H_{1}^{\prime}=H_{FA}-H_{0}^{\prime}$ which is the
difference between the original FA-Hamiltonian and $H_{0}^{\prime}$, has zero
energy expectation value in the FAIR ground state. In addition it is shown
in\ second order self-consistent perturbation theory (see appendix C) that the
total occupation of all perturbation states is only of the order of $10^{-4}$,
i.e. the FAIR ground state has still an amplitude of $0.9998$. This is
important when an electron or hole is injected into the ground state.

The excitation spectrum is obtained by injecting an electron or a hole into
the ground state. The resulting excited states interact via the perturbation
Hamiltonian $H_{1}^{\prime}$ and yield a spectrum of energy resonances. It
turns out that the injection of an electron into the spin-up conduction band
creates also transition between the spin-down conduction band and the nest states.

The resulting density of states possesses the shape of a resonance curve.
However, the resonance width is about twice the value of the mean-field
theory. As a consequence the height of the resonance density of states is
reduced by a factor of two. Since the formation of a magnetic moment depends
on the product of the Coulomb interaction and the density of d-states, one
would expect that the mean field overestimates the tendency towards a magnetic
moment. Indeed I observed in the first paper about the magnetic ground state
that the formation of a magnetic moment requires about twice the Coulomb
energy that the mean field theory predicts. This consorts well with the
present finding of the reduced resonance density of states.

The next step in the future investigation is the calculation of the density of
states of the Kondo resonance within the FAIR model. For this calculation one
has to use a Wilson spectrum with a logarithmic energy scale.%

\[
\]
\newpage

\appendix{}

\section{The FAIR approach}

\subsection{The Friedel impurity}

The basic idea of the FAIR method can be best explained for a Friedel
resonance with the Hamiltonian
\[
H_{F}=\sum_{\nu=0}^{N-1}\varepsilon_{\nu}c_{\nu}^{\dag}c_{\nu}+E_{d}d^{\dag
}d+\sum_{\nu=0}^{N-1}V_{\nu}^{sd}[d^{\dag}c_{\nu}+c_{\nu}^{\dag}d]
\]
This is done in the following steps:

\begin{enumerate}
\item from the free (or s-) electron basis an artificial Friedel resonance
(\textbf{FAIR)} state is constructed with $a_{0}^{\dag}=%
%TCIMACRO{\tsum _{\nu}}%
%BeginExpansion
{\textstyle\sum_{\nu}}
%EndExpansion
\alpha_{0}^{\nu}c_{\nu}^{\dag}$ together with a full orthonormal basis
$\left\{  a_{i}^{\dag}\right\}  $ so that the free electron-electron
Hamiltonian $H_{fe}=\sum_{\nu=0}^{N-1}\varepsilon_{\nu}c_{\nu}^{\dag}c_{\nu}$
takes the form
\[
H_{fe}=\sum_{i=1}^{N-1}E_{i}^{\left(  a\right)  }a_{i}^{\dag}a_{i}%
+E_{0}^{\left(  a\right)  }a_{0}^{\dag}a_{0}+\sum_{i=1}^{N-1}V_{i}^{\left(
a\right)  fr}\left[  a_{0}^{\dag}a_{i}+a_{i}^{\dag}a_{0}\right]
\]
The requirement that the matrix elements between different $a_{j}^{\dag}$ and
$a_{k}^{\dag}$ $\left(  i,k\neq0\right)  $ vanish has the consequence that a
given FAIR state $a_{0}^{\dag}$ determines uniquely the full basis $\left\{
a_{i}^{\dag}\right\}  $. In the new basis the total Friedel Hamiltonian takes
the form%
\[
H_{F}=H_{0}^{\prime}+H_{1}^{\prime}%
\]
where
\begin{equation}
H_{0}^{\prime}=\sum_{i=1}^{N-1}E_{i}^{\left(  a\right)  }a_{i}^{\dag}%
a_{i}+E_{0}^{\left(  a\right)  }a_{0}^{\dag}a_{0}+E_{d}d^{\dag}d+V_{0}%
^{\left(  a\right)  sd}[d^{\dag}a_{0}+a_{0}^{\dag}d]
\end{equation}
The perturbation Hamiltonian has the form%
\begin{equation}
H_{1}^{\prime}=\left\{  \sum_{i=1}^{N-1}V_{i}^{\left(  a\right)  fr}\left[
a_{0}^{\dag}a_{i}+a_{i}^{\dag}a_{0}\right]  +\sum_{i=1}^{N-1}V_{i}^{\left(
a\right)  sd}\left[  d^{\dag}a_{i}+a_{i}^{\dag}d\right]  \right\}
\end{equation}
Here the new matrix elements are given as $V_{i}^{\left(  a\right)  fr}$ and
$V_{i}^{\left(  a\right)  sd}$ .

\item A trial state $\Psi_{F}$ is defined as%
\begin{equation}
\Psi_{F}=\left[  A_{0}a_{0}^{\dag}+A_{d}d^{\dag}\right]  \left\vert
\mathbf{0}_{a}\right\rangle \label{Psi_F}%
\end{equation}
where $\left\vert \mathbf{0}_{a}\right\rangle =\prod_{j=1}^{n-1}a_{j}^{\dag
}\left\vert \Phi_{0}\right\rangle ,$ $n=N/2$. The right side is abbreviated as%
\[
\Psi_{F}=%
%TCIMACRO{\tsum _{\alpha=0,d}}%
%BeginExpansion
{\textstyle\sum_{\alpha=0,d}}
%EndExpansion
A_{\alpha,\beta}\Psi_{\alpha}%
\]
where, for example, for $\alpha=d$ one has $\Psi_{d}=d^{\dag}\left\vert
\mathbf{0}_{a}\right\rangle $

\item The energy expectation value $E_{00}=\left\langle \Psi_{F}\left\vert
H_{F}\right\vert \Psi_{F}\right\rangle $ of the $H_{0}^{\prime}$ with respect
to the trial state $\Psi_{F}$ is calculated. The contribution of
$H_{1}^{\prime}$ is not included. This results in a $2x2$ secular matrix
\[
\left(
\begin{array}
[c]{cc}%
%TCIMACRO{\tsum _{i=1}^{n-1}}%
%BeginExpansion
{\textstyle\sum_{i=1}^{n-1}}
%EndExpansion
E_{i}+E_{0}^{\left(  a\right)  } & V_{0}^{\left(  a\right)  sd}\\
V_{0}^{\left(  a\right)  sd} &
%TCIMACRO{\tsum _{i=1}^{n-1}}%
%BeginExpansion
{\textstyle\sum_{i=1}^{n-1}}
%EndExpansion
E_{i}+E_{d}%
\end{array}
\right)
\]
whose lowest eigenvalue yields $E_{00}$ and the corresponding eigenvector
yields the coefficients $A_{0},A_{d}$.

\item The FAIR state is rotated (by variation) in the $N$-dimensional Hilbert
space until the lowest eigenvalue of the secular matrix reaches a minimum.
\end{enumerate}

It has been shown by the author \cite{B91}, \cite{B92} that this procedure
results in the \textbf{exact} $n$-particle ground state of a Friedel
Hamiltonian (given by (\ref{Psi_F})). The matrix elements of the perturbation
Hamiltonian $H_{1}^{\prime}$ between this ground state and any excited state vanish.

It is interesting to look at the result in some more detail.

\begin{itemize}
\item The states $a_{i}^{\dag}$ $\left(  i\neq0\right)  $ enter the secular
matrix only through the total energy of the occupied states $%
%TCIMACRO{\tsum _{i=1}^{n-1}}%
%BeginExpansion
{\textstyle\sum_{i=1}^{n-1}}
%EndExpansion
E_{i}$ and contribute only the the background energy.

\item The coefficients and the relative weight of the states $a_{0}^{\dag}$
and $d^{\dag}$ are only determined by the energies of the FAIR and the d-state
$E_{0}^{\left(  a\right)  },E_{d}$ and their coupling $V_{0}^{\left(
a\right)  sd}$.
\end{itemize}

In a way one can say that the states $a_{i}^{\dag}$ prepare just the
background - a kind of nest - for $a_{0}^{\dag}$ and $d^{\dag}$. The secular
matrix represents an effective Hamiltonian for these two states in the nest.
In the following I will call the secular matrix without the kinetic energy $%
%TCIMACRO{\tsum _{i=1}^{n-1}}%
%BeginExpansion
{\textstyle\sum_{i=1}^{n-1}}
%EndExpansion
E_{i}$ the \textbf{nest Hamiltonian}.%
\[
H^{nst}=\left(
\begin{array}
[c]{cc}%
E_{0}^{\left(  a\right)  } & V_{0}^{\left(  a\right)  sd}\\
V_{0}^{\left(  a\right)  sd} & E_{d}%
\end{array}
\right)
\]

The state $a_{0}^{\dag}$ represents an artificially inserted Friedel resonance
state. Therefore I call $a_{0}^{\dag}$ a "\textbf{F}riedel \textbf{A}%
rtificially \textbf{I}nserted \textbf{R}esonance" state or \textbf{FAIR}%
-state. The use of the FAIR-states is at the heart of my approach to the FA-
and Kondo impurity problem. Therefore I call this approach the \textbf{FAIR }method.

\subsection{From mean field to the FAIR magnetic state}

The Hamiltonian of the Friedel-Anderson impurity is given in equ.
(\ref{H_FA}). One obtains the mean-field Hamiltonian from equ.(\ref{hfa0}) by
replacing $n_{d\uparrow}n_{d\downarrow}$ =%
%TCIMACRO{\TEXTsymbol{>}}%
%BeginExpansion
$>$%
%EndExpansion
$n_{d\uparrow}\left\langle n_{d\downarrow}\right\rangle $ $+\left\langle
n_{d\uparrow}\right\rangle n_{d\downarrow}$ $-\left\langle n_{d\uparrow
}\right\rangle \left\langle n_{d\downarrow}\right\rangle $. After adjusting
$\left\langle n_{d\uparrow}\right\rangle $ and $\left\langle n_{d\downarrow
}\right\rangle $ self-consistently one obtains two Friedel resonance
Hamiltonians with a spin-dependent energy of the $d_{\sigma}$-state:
$E_{d,\sigma}$ $=E_{d}+U\left\langle n_{d,-\sigma}\right\rangle $.
\[
H_{mf}=%
%TCIMACRO{\tsum _{\sigma}}%
%BeginExpansion
{\textstyle\sum_{\sigma}}
%EndExpansion
\left\{  \sum_{\nu=1}^{N}\varepsilon_{\nu}c_{\nu\sigma}^{\dag}c_{\nu\sigma
}+E_{d\sigma}d_{\sigma}^{\dag}d_{\sigma}+\sum_{\nu=1}^{N}V_{sd}(\nu
)[d_{\sigma}^{\dag}c_{\nu\sigma}+c_{\nu\sigma}^{\dag}d_{\sigma}]\right\}
\]
The mean-field wave function is a product of two Friedel ground states for
spin up and down $\Psi_{mf}=\Psi_{F\uparrow}\Psi_{F\downarrow}$ .

Now we express each Friedel ground state $\Psi_{F\sigma}$ by the FAIR
solution, for example%
\[
\Psi_{F,\uparrow}=\left(  A_{a,\uparrow}a_{0,\uparrow}^{\dag}+A_{d,\uparrow
}d_{\uparrow}^{\dag}\right)
%TCIMACRO{\tprod \limits_{i=1}^{n-1}}%
%BeginExpansion
{\textstyle\prod\limits_{i=1}^{n-1}}
%EndExpansion
a_{i,\uparrow}^{\dag}\Phi_{0}%
\]

For the two Friedel states in the mean-field wave function I use the form of
equ. (\ref{Psi_F}) and obtain for the mean-field solution
\begin{align}
\Psi_{mf}  &  =\left[  \left(  A_{a,\uparrow}a_{0,\uparrow}^{\dag
}+A_{d,\uparrow}d_{\uparrow}^{\dag}\right)  \prod_{i=1}^{n-1}a_{i\uparrow
}^{\dag}\right]  \left[  \left(  A_{b,\downarrow}a_{0-\downarrow}^{\dag
}+A_{s,\downarrow}d_{\downarrow}^{\dag}\right)  \prod_{i=1}^{n-1}%
b_{i\downarrow}^{\dag}\right]  \Phi_{0}\label{Psi_mf}\\
&  =\left[  A_{a,b}a_{0\uparrow}^{\dag}b_{0\downarrow}^{\dag}+A_{a,d}%
a_{0\uparrow}^{\dag}d_{\downarrow}^{\dag}+A_{d,b}d_{\uparrow}^{\dag
}b_{0\downarrow}^{\dag}+A_{d,d}d_{\uparrow}^{\dag}d_{\downarrow}^{\dag
}\right]  \left\vert \mathbf{0}_{a,\uparrow}\mathbf{0}_{b,\downarrow
}\right\rangle \nonumber
\end{align}
where $\left\{  a_{i,\uparrow}^{\dag}\right\}  $ and $\left\{  b_{i,\downarrow
}^{\dag}\right\}  $ are two (different) bases of the $N$-dimensional Hilbert
space. This solution can be rewritten as equation (\ref{Psi_MS}).

In the mean-field solution $\Psi_{mf}$ the coefficients $A_{\alpha,\beta}$ are
restricted by two conditions $A_{a,\uparrow}^{2}+A_{d,\uparrow}^{2}=1$
($A_{b,\downarrow}^{2}+A_{d,\downarrow}^{2}=1$). Therefore this state does not
describe well the correlation effects.

In contrast the state (\ref{Psi_MS}) opens a wide playing field for improving
the solution: (i) The FAIR states $a_{0}^{\dag}$ and $b_{0}^{\dag}$ can be
individually optimized, each one defining a whole basis $\left\{  a_{i}^{\dag
}\right\}  $ and $\left\{  b_{j}^{\dag}\right\}  $. This yields a much better
treatment of the correlation effects. The resulting state is denoted as the
(potentially) magnetic state $\Psi_{MS}$. The magnetic state $\Psi_{MS}$ has
the same structure as the mean field solution $\Psi_{mf}$; the only difference
is that its components are optimized for the Friedel-Anderson Hamiltonian.

\section{Self-consistent Perturbation}

In the construction of the magnetic ground state $\Psi_{MS}$ only the
Hamiltonian $H_{0}^{\prime}$ (equ. \ref{H_0'}) has been used. The expectation
value of the "perturbation" Hamiltonian $H_{1}^{\prime}$ is zero,
$\left\langle \Psi_{MS}\left\vert H_{1}^{\prime}\right\vert \Psi
_{MS}\right\rangle =0$ where $H_{1}^{\prime}=H_{1\uparrow}+H_{1\downarrow}$
and
\begin{equation}
H_{1,\sigma}^{\prime}=\sum_{j=1}^{N-1}V_{j}^{\left(  a\right)  fr}\left[
a_{0,\sigma}^{\dag}a_{j,\sigma}+a_{j,\sigma}^{\dag}a_{0,\sigma}\right]
+\sum_{j=1}^{N-1}V_{j}^{\left(  a\right)  sd}\left[  d_{\sigma}^{\dag
}a_{j,\sigma}+a_{j,\sigma}^{\dag}d_{\sigma}\right]  \label{H1p}%
\end{equation}

But $H_{1}^{\prime}$ yields transitions from the ground state $\Psi_{MS}$ into
excited states. From equ. (\ref{H1p}) one recognizes that $H_{1\uparrow
}^{\prime}$ only permits transitions between the nest $\left(  d_{\uparrow
}^{\dag},a_{0\uparrow}^{\dag}\right)  $ and a band state $a_{j\uparrow}^{\dag
}$ but no transition among band states.

If one considers only final states through this transitions which are linear
in $H_{1}^{\prime}$ then Fig.9 shows the possible final states. For the
spin-up band this are the transitions from the nest state into an electron
excitation $a_{j}^{\dag}$, leaving the spin-up part of the nest empty or a
transition from an occupied state by creating a hole $a_{j}$ and filling both
state $a_{0\uparrow}^{\dag}$ and $d_{\uparrow}^{\dag}$ of the spin-up part of
the nest. The corresponding states are generated for the spin-down band.%

\begin{align*}
&
%TCIMACRO{\FRAME{itbpF}{5.335in}{1.8198in}{0in}{}{}{corl181_{2}f.eps}%
%width 5.335in;  height 1.8198in;  depth 0in;  original-width 5.0668in;
%original-height 1.7111in;  cropleft "0";  croptop "1";  cropright "1";
%cropbottom "0";  filename 'Fig9.eps';file-properties "XNPEU";}}}%
%BeginExpansion
{\includegraphics[
height=1.8198in,
width=5.335in
]%
{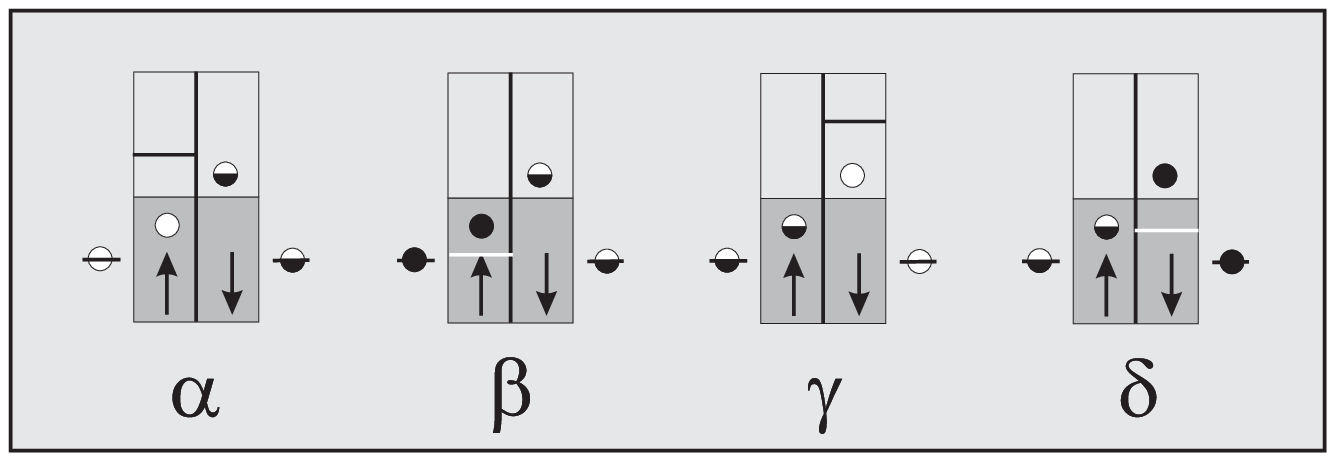}%
}%
%EndExpansion
\\
&
\begin{tabular}
[c]{l}%
Fig.9: Final states which can be obtained by transitions from the\\
ground state $\Psi_{MS}$ through the perturbation Hamiltonian $H_{1}^{\prime
}.$%
\end{tabular}
\end{align*}

As an example we consider the transitions in the spin-up band. Applying
$H_{1\uparrow}^{\prime}$ to the ground state $\Psi_{MS}$ yields%
\[
H_{1\uparrow}^{\prime}\Psi_{MS}=\sum_{j=1}^{N-1}\left\{
\begin{array}
[c]{c}%
\left[  V_{j}^{\left(  a\right)  fr}A_{a,b}+V_{j}^{\left(  a\right)
sd}A_{d,b}\right]  a_{j,\sigma}^{\dag}b_{0\downarrow}^{\dag}\\
+\left[  V_{j}^{\left(  a\right)  fr}A_{a,d}+V_{j}^{\left(  a\right)
sd}A_{d,d}\right]  a_{j,\sigma}^{\dag}d_{\downarrow}^{\dag}\\
+\left(  -V_{j}^{\left(  a\right)  fr}A_{d,b}+V_{j}^{\left(  a\right)
sd}A_{a,b}\right)  a_{j,\uparrow}a_{0\uparrow}^{\dag}d_{\uparrow}^{\dag
}b_{0\downarrow}^{\dag}\\
+\left(  -V_{j}^{\left(  a\right)  fr}A_{d,d}+V_{j}^{\left(  a\right)
sd}A_{a,d}\right)  a_{j,\uparrow}a_{0\uparrow}^{\dag}d_{\uparrow}^{\dag
}d_{\downarrow}^{\dag}%
\end{array}
\right\}  \left\vert \mathbf{0}_{a,\uparrow}\mathbf{0}_{b,\downarrow
}\right\rangle
\]

The top two lines represent an electron excitation, shown as the left state in
Fig.9 and the bottom two lines a hole excitation shown as the second state in
Fig.9. The final states and the corresponding matrix elements are collected in
table III.%

\begin{align*}
&
\begin{tabular}
[c]{|l|l|l|l|}\hline
$\mathbf{\Psi}_{i}$ & $\mathbf{\Psi}_{f}$ & $\left\langle \Psi_{i}\left\vert
H_{1}^{\prime}\right\vert \Psi_{f}\right\rangle $ & \textbf{excitation}%
\\\hline
$\Psi_{MS}$ & $a_{j\uparrow}^{\dag}b_{0\downarrow}^{\dag}\left\vert
\mathbf{0}_{a,\uparrow}\mathbf{0}_{b,\downarrow}\right\rangle $ & $\left(
V_{j}^{\left(  a\right)  fr}A_{a,b}+V_{j}^{\left(  a\right)  sd}%
A_{d,b}\right)  $ & electron, $j\geq n$\\\hline
$\Psi_{MS}$ & $a_{j,\uparrow}^{\dag}d_{\downarrow}^{\dag}\left\vert
\mathbf{0}_{a,\uparrow}\mathbf{0}_{b,\downarrow}\right\rangle $ & $\left(
V_{j}^{\left(  a\right)  fr}A_{a,d}+V_{j}^{\left(  a\right)  sd}%
A_{d,d}\right)  $ & electron, $j\geq n$\\\hline
$\Psi_{MS}$ & $a_{j,\uparrow}a_{0\uparrow}^{\dag}d_{\uparrow}^{\dag
}b_{0\downarrow}^{\dag}\left\vert \mathbf{0}_{a,\uparrow}\mathbf{0}%
_{b,\downarrow}\right\rangle $ & $\left(  -V_{j}^{\left(  a\right)  fr}%
A_{d,b}+V_{j}^{\left(  a\right)  sd}A_{a,b}\right)  $ & hole, $0<j<n$\\\hline
$\Psi_{MS}$ & $a_{j,\uparrow}a_{0\uparrow}^{\dag}d_{\uparrow}^{\dag
}d_{\downarrow}^{\dag}\left\vert \mathbf{0}_{a,\uparrow}\mathbf{0}%
_{b,\downarrow}\right\rangle $ & $\left(  -V_{j}^{\left(  a\right)  fr}%
A_{d,d}+V_{j}^{\left(  a\right)  sd}A_{a,d}\right)  $ & hole, $0<j<n$\\\hline
\end{tabular}
\ \ \ \ \\
&
\begin{tabular}
[c]{l}%
Table III: For the spin-up band the final states $\Psi_{f}$, and the matrix
elements\\
of $H_{1}^{\prime}$ between the ground state $\Psi_{MS}$ and the final states
$\Psi_{f}$ are listed.\\
The final states still have to be expanded into new eigenstates of the nest\\
plus band. For the spin-down band one obtains equivalent transitions.
\end{tabular}
\ \
\end{align*}

There are two important aspects of this result, (i) there are always two
transitions into each final state, for example $a_{0\uparrow}^{\dag
}\rightarrow a_{j\uparrow}^{\dag}$ and $d_{\uparrow}^{\dag}\rightarrow
a_{j\uparrow}^{\dag}$. These two transitions interfere and cancel each other
almost completely (as will be shown below). (ii) The first two final states in
table III (as well as the third and forth final state) are not eigenstates of
$H_{0}^{\prime}$. One has to expand these final states in terms of the
eigenstates of the nest, for example%
\[%
\begin{tabular}
[c]{|l|}\hline
$a_{j,\uparrow}a_{0\uparrow}^{\dag}d_{\uparrow}^{\dag}b_{0\downarrow}^{\dag
}\left\vert \mathbf{0}_{a,\uparrow}\mathbf{0}_{b,\downarrow}\right\rangle
$\\\hline
$a_{j,\uparrow}a_{0\uparrow}^{\dag}d_{\uparrow}^{\dag}d_{\downarrow}^{\dag
}\left\vert \mathbf{0}_{a,\uparrow}\mathbf{0}_{b,\downarrow}\right\rangle
$\\\hline
\end{tabular}
\ \ \ \ \ \Longleftrightarrow%
\begin{array}
[c]{c}%
a_{j,\uparrow}\Psi_{2/1}^{\left(  1\right)  }\\
a_{j,\uparrow}\Psi_{2/1}^{\left(  2\right)  }%
\end{array}
\]
This is in complete analogy to the calculation of the excitations in section III.

In Fig.10 the logarithm of the effective matrix element is plotted for the
first transition in table III. This represents an electron excitation which is
restricted to positive energies. As one recognizes the effective matrix
elements are strongly reduced in the energy range in which transitions are
possible. The value of $V_{eff}$ lies in the range between $10^{-3}$ and
$10^{-4}$ while the original matrix elements for $V_{sd}$ are $0.025$. This
applies for all eight possible excitations. In the energy range where an
excitation is permitted the matrix elements are strongly reduced. Among the
eight possible transitions there is only one transition whose matrix elements
exceed $10^{-3}$. This is the second one in table III where the matrix element
reaches values of $2\times10^{-3}$, still much smaller than the original
matrix elements.%

\begin{align*}
&
%TCIMACRO{\FRAME{itbpF}{3.9618in}{3.2312in}{0pt}{}{}{org181_{8}a.eps}%
%width 3.9618in;  height 3.2312in;  depth 0pt;  original-width 3.7559in;
%original-height 3.0585in;  cropleft "0";  croptop "1";  cropright "1";
%cropbottom "0";  filename 'Fig10.eps';file-properties "XNPEU";}}}%
%BeginExpansion
\raisebox{-0pt}{\includegraphics[
height=3.2312in,
width=3.9618in
]%
{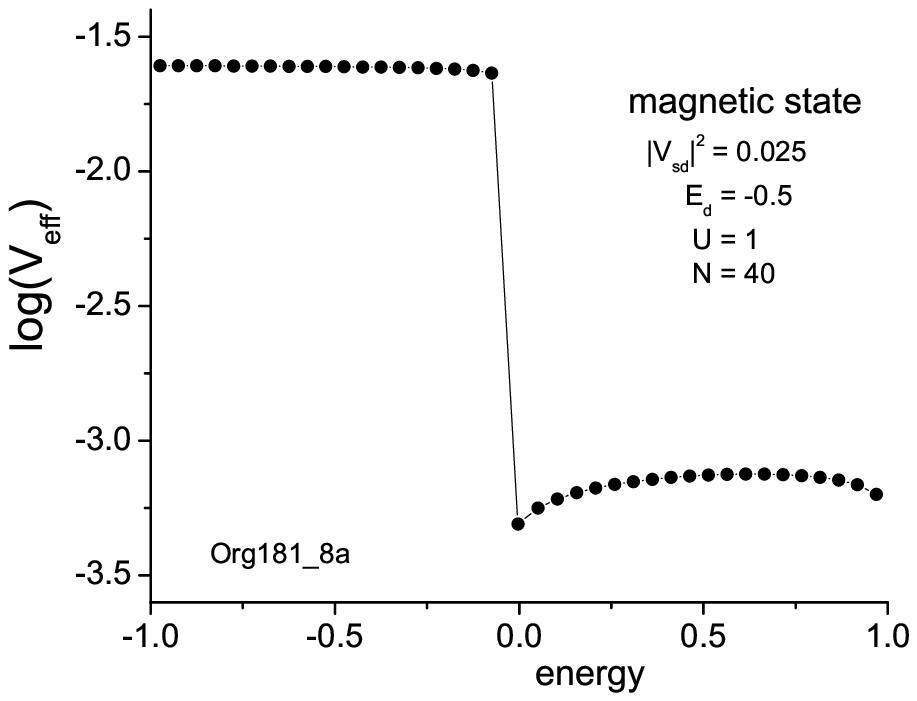}%
}%
%EndExpansion
\\
&
\begin{tabular}
[c]{l}%
Fig.10: The logarithm of the effective matrix elements for a transition\\
from the ground state $\Psi_{MS}$ into the excited state $a_{j\uparrow}^{\dag
}b_{0\downarrow}^{\dag}\left\vert \mathbf{0}_{a,\uparrow}\mathbf{0}%
_{b,\downarrow}\right\rangle $. This\\
transition is only possible for positive energies, and there the\\
interference between the the d-state $d_{\uparrow}^{\dag}$ and the FAIR state
$a_{0\uparrow}^{\dag}$\\
cancels the transition almost completely.
\end{tabular}
\end{align*}

The strong reduction of the matrix elements is due to the introduction of the
FAIR states which compensate the transitions involving the d-states. It is
also the reason why the magnetic ground state is so well represented by
$\Psi_{MS}$.

With the eigenenergies of the states in Fig.9 and the matrix elements for the
transition from the ground state $\Psi_{MS}$ into these states one can now
perform a self-consistent perturbation calculation. The number of excited
states is $2\left(  N-1\right)  $ for each spin-band. One can build the the
full secular matrix for these states which consists of $4\left(  N-1\right)
+1$ states, where the additional one is the ground state. The result for our
standard example ($E_{d}=-0.5,$ $U=1$, $\left\vert V_{sd}^{0}\right\vert
^{2}=0.025$, $N=40)$ is rather dramatic. The weight of all $4\left(
N-1\right)  $ excitations together is only about $10^{-4}$. The amplitude of
$\Psi_{MS}$ after diagonalization is $0.99985$. This demonstrates the FAIR
solution is an excellent approximation to the ground state.

\section{Green's functions}

With the secular Hamiltonian $H^{xct}$ in section 3.1 one can construct the
Green's functions (GF)of the excitations $\varphi_{\nu}$%
\[%
%TCIMACRO{\tsum _{\nu}}%
%BeginExpansion
{\textstyle\sum_{\nu}}
%EndExpansion
\left(  \varepsilon+is-H^{xct}\right)  _{\mu,\nu}G_{\nu,\kappa}=\delta
_{\mu,\kappa}%
\]
or
\[
\mathbf{G=}\left(  \varepsilon+is-\mathbf{H}^{xct}\right)  ^{-1}%
\]
The resulting diagonal elements of $\mathbf{G}$ are the Green's function of
the excitations, for example $G_{11}\left(  \varepsilon\right)  $ is the
Green's function of $\Psi_{2/1}^{\left(  1\right)  }$. Since the $G_{\mu,\mu
}\left(  \varepsilon\right)  $ are a function of the energy the above relation
is not very practical for a numerical nor analytical calculation of
$G_{\mu,\mu}\left(  \varepsilon\right)  $. For a numerical evaluation it is
much easier to calculate the eigenvalues $E_{\mu}^{xct}$ and eigenvectors
$\psi_{\mu}$ of $H^{xct}$ where $\psi_{\mu}=%
%TCIMACRO{\tsum _{\nu}}%
%BeginExpansion
{\textstyle\sum_{\nu}}
%EndExpansion
\psi_{\mu}^{\nu}\varphi_{\nu}$. Then the Green's function $G_{\mu,\mu}$ of the
excitation can be expressed as
\[
G_{\nu,\nu}\left(  \varepsilon\right)  =%
%TCIMACRO{\tsum _{\nu}}%
%BeginExpansion
{\textstyle\sum_{\nu}}
%EndExpansion
\frac{\left\vert \psi_{\mu}^{\nu}\right\vert ^{2}}{\varepsilon-E_{\mu}%
^{xct}+is}%
\]
Each energy eigenvalue contributes to the spectrum of the state $\varphi_{\nu
}.$

For the particle density of states we need the Green's function (or the
spectrum) of the $\frac{N}{2}$ states $a_{j\uparrow}^{\dag}\Psi_{MS}$ and
$a_{0\uparrow}^{\dag}\Psi_{MS}$ and $d_{\uparrow}^{\dag}\Psi_{MS}$. The $N/2$
electron Green's functions $G_{a_{j},a_{j}}\left(  \varepsilon\right)  $
follow directly from the above calculation. The $G_{a_{0},a_{0}}\left(
\varepsilon\right)  $ and $G_{d,d}\left(  \varepsilon\right)  $ for spin up
have to be derived from the Green's functions of $\Psi_{2/1}^{\left(
1\right)  }$ and $\Psi_{2/1}^{\left(  2\right)  }$.

In Fig.3 a spin-up electron is injected into the nest. This electron can be
injected into the $a_{0\uparrow}^{\dag}$ or the $d_{\uparrow}^{\dag}$ state.
These processes yield complementary amplitudes of
\[%
\begin{tabular}
[c]{lll}%
$a_{0\uparrow}^{\dag}\Psi_{MS}=\left(  A_{d,b}a_{0}^{\dag}d_{\uparrow}^{\dag
}b_{0\downarrow}^{\dag}+A_{d,d}a_{0}^{\dag}d_{\uparrow}^{\dag}d_{\downarrow
}^{\dag}\right)  \left\vert \mathbf{0}_{a,\uparrow}\mathbf{0}_{b,\downarrow
}\right\rangle $ & for & $a_{0}^{\dag}$\\
$d_{\uparrow}^{\dag}\Psi_{MS}=-\left(  A_{a,b}a_{0\uparrow}^{\dag}d_{\uparrow
}^{\dag}b_{0\downarrow}^{\dag}+A_{a,d}a_{0\uparrow}^{\dag}d_{\uparrow}^{\dag
}d_{\downarrow}^{\dag}\right)  \left\vert \mathbf{0}_{a,\uparrow}%
\mathbf{0}_{b,\downarrow}\right\rangle $ & for & $d_{\uparrow}^{\dag}$%
\end{tabular}
\ \ \
\]

Each of the two resulting states has to be expanded in the two eigenstates
$\Psi_{2/1}^{\left(  1\right)  }$ and $\Psi_{2/1}^{\left(  2\right)  }$ of the
nest. For example, the amplitudes for $a_{0\uparrow}^{\dag}\Psi_{MS}$ are
obtained through the scalar product between $a_{0\uparrow}^{\dag}\Psi_{MS}$
and $\Psi_{2/1}^{\left(  \alpha\right)  }$. Then they can make a transition
into any of the other states with a single excitation, for example the state
$a_{j\uparrow}^{\dag}\Psi_{MS}$. Again the resulting amplitudes interfere. So
the particle components of the GFs $G_{a_{0},a_{0}}\left(  \varepsilon\right)
$ and $G_{d,d}\left(  \varepsilon\right)  $ are composed of the GFs of
$\Psi_{2/1}^{\left(  \alpha\right)  }$ with a weight which is the square of
the scalar \ product between the amplitude in equ. (\ref{a_Psi_MS}) and
$\Psi_{2/1}^{\left(  \alpha\right)  }$. These GFs $G_{a_{0},a_{0}}\left(
\varepsilon\right)  $ and $G_{d,d}\left(  \varepsilon\right)  $ have also a
hole component.

\end{document}